\documentclass[lineno]{JFM-FLM_Au}

\usepackage{cuted}
\usepackage{amsmath}
\usepackage{fdsymbol}
\usepackage[utf8]{inputenc}
\usepackage{longtable}
\usepackage{wrapfig}
\usepackage{balance}

\lefttitle{J. Galen Wang, Umesh Dhumal, Monica E. A. Zakhari and Roseanna N. Zia}
\righttitle{Journal Name XXXX}

\title{Frenkel’s entropy-exchange mechanism in monodisperse, nearly hard-sphere colloids: minimal perturbations to access fluid–crystal coexistence}

\author{J. Galen Wang\aff{1}, Umesh Dhumal\aff{1}, Monica E. A. Zakhari\aff{2} \and Roseanna N. Zia\aff{1}}

\affiliation{\aff{1}Mechanical and Aerospace Engineering, University of Missouri, 416 S. Sixth St., Columbia, Missouri 65201, USA
\aff{2}Mechanical Engineering, Eindhoven University of Technology, Gemini-Zuid, Eindhoven, The Netherlands}

\corresau{Roseanna N. Zia: rzia@missouri.edu}

\begin{document}
\maketitle
\begin{abstract}
Entropically driven fluid--solid transitions in monodisperse, purely repulsive hard spheres (MPRHS) are well established in theory, simulation, and experiment for atomic and colloidal systems. For MPRHS, however, coexistence is usually located via bulk free-energy calculations; the underlying microscopic balance between configurational and vibrational entropy is left implicit. Frenkel clarified this mechanism explicitly as an exchange of long-range configurational entropy for short-range vibrational entropy, but in the pristine MPRHS limit the nucleation barrier near coexistence is so high that phase separation is predicted only on astronomical time scales. Consistent with this, even unbiased simulations do not show spontaneous, equilibrium fluid--crystal coexistence; transient mixtures are mostly overtaken by a single phase; observed coexistence is still algorithmically-driven. Nearly hard-sphere colloid {\em experiments} do observe fluid--crystal coexistence, but always in the presence of unavoidable triggers such as gravity, walls, and polydispersity. We treat the hard-sphere phase diagram as settled and ask how the entropic exchange mechanism can be revealed in nearly hard-sphere colloidal simulations. We probe the mechanism on finite time scales by introducing minimal perturbations that trigger phase separation: small reductions in hardness that increase locally accessible free volume (and thus gently increase vibrational entropy), and 2--4\% distributed crystal seeds.  These perturbations produce coexisting fluid and crystal domains with crystal fraction, phase envelope and osmotic pressure that, with systematically increasing particle hardness, approach the hard-sphere limit. These results demonstrate that slight enhancements to vibrational entropy provide a dynamically accessible route to realizing the long-range/short-range entropy exchange required for phase separation.

\end{abstract}

\section{Introduction}
\label{sec:intro}
Phase transitions in atomic and colloidal systems arise from competition among contributions to the free energy. Purely entropic transitions have been predicted and observed in hard-particle systems by theory, experiments, and simulations for nearly a century. In systems with shape anisotropy or size polydispersity, each distinct orientation or size provides an entropic drive that competes with the others, and this competition produces phase separation. For example, Onsager showed that shape anisotropy can generate configurational entropy that drives liquid-crystalline order in highly anisometric particles \citep{onsager1949effects}, a mechanism later framed by Frenkel as an exchange between orientational and translational entropy that underlies, e.g., the isotropic-nematic transition \citep{frenkel2000perspective}. Subsequent work mapped phase envelopes for atomic and colloidal systems and quantified how phase behavior is reshaped by \textbf{softness} \citep{hoover1971thermodynamic, Robbins1988, Meijer1991, piazza1993equilibrium, Lowen1993, Lowen1993b, nemeth1995solid, senff1999temperature, likos2001effective, castelletto2002liquid, Laurati2005, archer2005density, likos2006soft, mladek2007phase, mladek2007clustering, Vlassopoulos2014, gupta2015dynamic, pelaez2015impact, Zakhari2017, erigi2023phase}, \textbf{shape anisotropy} \citep{eppenga1984monte, camp1997phase, cuetos2007kinetic, cinacchi2010phase, miller2010crystallization, kallus2011dense, agarwal2011mesophase, haji2011phase, jiao2011communication, avendano2012phase, marechal2012freezing, peroukidis2013phase, dijkstra2014entropy, boles2016self, karas2019phase, lim2023engineering}, and \textbf{size polydispersity} (via fractionation and polycrystallinity) \citep{kranendonk1991computer, bartlett1992superlattice, eldridge1993entropy, han1994freezing, dijkstra1998phase, dijkstra1999direct, bw-99, fs-03, zubarev2005condensation, zaccarelli2009crystallization, wilding2010phase, hopkins2011phase, filion2011self, dijkstra2014entropy, boles2016self, koshoji2021diverse, koshoji2021densest}. Two recent reviews summarize this landscape across theory, experiment, and simulation \citep{royall2024colloidal, wangInreviewElusive}. Across these studies, two hallmarks of first-order transitions consistently appear: {phase} \textit{transition} between pure states and {phase} \textit{separation} into coexisting domains.

Systems of monodisperse, purely repulsive hard spheres (MPRHS) also undergo fluid-crystal transitions. While the distinct contributors to phase transition are less obvious (only a single size of isotropic particles), prediction of MPRHS phase behavior in colloids dates back to predictions for atomic systems from Kirkwood and Monroe, who predicted the melting point of atomic hard spheres in 1941 \citep{kirkwood1941statistical}; Alder and Wainwright's simulations (1957--1960), traced the fluid and solid lines \citep{alder1957phase,alder1959studies, alder1960studies}, establishing the \textit{transition}. Hoover and Ree later confirmed these lines and, using thermodynamic arguments, deduced the coexistence tie line and the freezing/melting volume fractions $\phi_F=0.494$ and $\phi_M=0.545$ that are standard in the hard-sphere literature \citep{hr-68}. Experiments, which necessarily approximate but cannot realize perfectly hard interactions, reproduced the full phase diagram in colloidal dispersions---most notably Pusey and van Megen's seminal study showing fluid, crystal, and explicit coexistence obeying a lever rule \citep{pvM-86} and the x‑ray stratification measurements of Russel and co‑workers \citep{prczcdo-96}. Thus, entropically driven first-order transitions in MPRHS are thoroughly established: distinct phases, phase envelopes, and $\phi_F$ and $\phi_M$ from theory, simulation, and experiment \citep{wangInreviewElusive}. 

Although the fluid–solid transition in MPRHS is long established by theory, its purely entropic origin was initially controversial because it seemed to imply that the more ordered crystal must somehow have higher entropy than the fluid \citep{uhlenbeck1963p, ackerson1993order, frenkel-93}. Classical liquid-state approaches effectively bypass having to make this microscopic competition explicit. They first compute separate equations of state for the homogeneous fluid and crystal branches using virial expansions and related frameworks \citep{thiele1963equation, wertheim1963exact, mcquarrie-76, ree1964fifth, ree1967seventh, cs-69, hall1972another, clisby2006ninth, schultz2014fifth}, and then locate coexistence by enforcing equality of pressure and chemical potential between the two branches \citep{frenkel1984new, wilding2000freezing, frenkel2002understanding, vega2007revisiting, odriozola2009replica, bannerman2010thermodynamic, nayhouse2011monte, fernandez2012equilibrium, statt2016monte, ustinov2017thermodynamics, pieprzyk2019thermodynamic, moir2021tethered}. Because virial and free-energy methods treat the fluid and crystal separately and assume analyticity within each phase, the first-order transition emerges from matching free energies across phases rather than from an explicit description of which microscopic entropic contributions are increasing or decreasing. In systems with attractions, polydispersity, or anisotropy, one can often appeal to additional energetic or entropic terms, but in pristine MPRHS the only driver is entropy, and because there is only a single size and no anisotropy, the underlying “mechanistic competition” is obscured. Frenkel proposed to make this explicit by viewing the transition as an exchange in which a loss of long-range configurational entropy is compensated by a gain in short-range vibrational entropy \citep{frenkel-93}. Making that configurational–vibrational entropy exchange operational is essential for building simulation models that do not just reproduce the hard-sphere phase diagram, but also interrogate how MPRHS actually undergo phase separation in practice \citep{wangInreviewElusive}.

However, decades of simulations that nominally match the atomic MPRHS model \emph{still} do not report spontaneously formed, long-lived fluid--crystal coexistence starting from an unbiased homogeneous state; explicit coexistence appears only when one introduces either algorithmic drivers or crystal-nucleating triggers~\citep{filion2010crystal,filion2011simulation,isobe2015hard,wangInreviewElusive}. To be clear, the first-order transition between fluid and crystal in monodisperse, purely repulsive hard spheres is firmly established by liquid-state theory, free-energy calculations, simulations, and experiments, and we take the hard-sphere equation of state and coexistence window as given~\citep{hr-68,cs-69,hall1972another,mcquarrie-76,kolafa2004accurate,pieprzyk2019thermodynamic,pvM-86,prczcdo-96}. Our aim is {\em not} to revisit or qualify that thermodynamic result. The issue is kinetic: in pristine MPRHS simulations (perfectly hard, strictly monodisperse spheres with no templates, walls, gravity, or softness) the equilibrium coexistence state is dynamically inaccessible on practical time scales. Unbiased EDMD or Brownian simulations prepared in the coexistence density interval typically remain as long-lived homogeneous fluid or homogeneous crystal, or else crystallize completely after a nucleation event, rather than developing a stationary two-phase coexistence state, consistent with nucleation-theory arguments that the barrier diverges at coexistence~\citep{filion2010crystal,filion2011simulation,fiorucci2020effect,wohler2022hard,pieprzyk2019thermodynamic}.

In addition to Brownian and event-driven molecular dynamics, Monte Carlo algorithms have been used to explore truly MPRHS phase behavior, done elegantly by ~\cite{isobe2015hard}, where coexistence is indeed realized---but with important caveats. They compare ECMC to LMC and EDMD, showing that ECMC very efficiently reproduces the hard-sphere equation of state and coexistence region, and that mixed fluid--crystal morphologies can be obtained within the coexistence window. Isobe and Krauth thus meet their intended goal: they convincingly demonstrate that different algorithms produce consistent equilibrium properties and that ECMC can map the hard-sphere EOS and coexistence region with high efficiency. That study is therefore an important \emph{thermodynamic} benchmark rather than a resolution of the \emph{kinetic} issue we address. Indeed, the authors explicitly note that simulation in the coexistence region remains difficult even with ECMC. To wit, coexistence states are always reached by using equilibrium-sampling Monte Carlo schemes (LMC/ECMC), whose moves are explicitly designed to accelerate equilibration and, at parts of the coexistence region, by also starting from crystal-rich initial configurations. Their observed behavior is entirely consistent with algorithmically driven equilibration: these schemes are constructed to reach the correct equilibrium mixture if one waits long enough, but they do so via nonphysical moves whose effective nucleation rates are set by the algorithm, not by the underlying colloidal dynamics. In this sense, the emergence of coexistence in their simulations always reflects algorithmically driven equilibration toward the known equilibrium mixture, rather than spontaneous colloidal dynamics under Brownian or Newtonian equations of motion. This behavior is consistent with the view that the coexistence state is thermodynamically well-defined but extremely difficult to realize dynamically in pristine MPRHS with physically motivated dynamics on accessible time scales.

Against this backdrop, it is useful to distinguish two broad ways in which simulations \emph{do} achieve fluid--crystal coexistence---but again, not spontaneously. One route uses equilibrium-sampling algorithms such as Metropolis Monte Carlo, event-chain Monte Carlo, and related enhanced-sampling schemes~\citep{frenkel1984new,vega2007revisiting,odriozola2009replica,fernandez2012equilibrium,statt2016monte,ustinov2017thermodynamics,isobe2015hard}. These methods are thermodynamically unbiased in the sense that they converge to the correct hard-sphere equilibrium ensemble, but they do so via algorithmically driven equilibration: moves that do not aim to reproduce Brownian or Newtonian trajectories are deliberately constructed to accelerate exploration of configuration space and the approach to equilibrium. As a result, any finite nucleation barrier will eventually be overcome, and the apparent ``nucleation rate'' is controlled by the move set, not by colloidal kinetics. A second route uses strong physical or geometric triggers---such as crystalline seeds, pre-constructed slabs (``direct coexistence''), walls, gravity, appreciable softness, or size polydispersity---which lower or bypass the metastable barrier and reliably generate coexisting domains~\citep{auer2001prediction,auer2001suppression,espinosa2013fluid,espinosa2014mold,hermes2011nucleation,tateno2019influence,espinosa2016seeding,montero2020young,montero2020interfacial,gispen2024finding,pvM-86,prczcdo-96,rutgers1996measurement,hw-09,rpw-13}. In this work, we take the hard-sphere phase diagram as settled, and focus instead on the kinetic bottleneck: we introduce deliberately minimal, quantitatively small perturbations---very slight deviations from perfect hardness and extremely weak, distributed crystalline seeds---while preparing the system in the coexistence region. These controlled deviations are chosen to leave the hard-sphere thermodynamics essentially unchanged, but to be just strong enough to break metastability on finite time scales, thereby rendering the phase-separation pathway observable. Our study is therefore aimed squarely at this kinetic question: starting from homogeneous, nearly MPRHS fluids evolved with Brownian dynamics, we ask what minimal, physically motivated perturbations to local vibrational entropy (finite softness) and, using very weak distributed seeding, are required to make the coexistence state actually appear and to expose Frenkel's configurational--vibrational entropy exchange mechanism in a system that is otherwise as close as possible to the ideal hard-sphere limit.

Frenkel's mechanism also sets expectations for ``pristine" simulations---perfectly hard, strictly monodisperse, repulsive spheres with system size governed by the Law of Large Numbers (LLN). In finite systems, unbiased simulations primarily report nucleation for $\phi>0.53$ and, crucially, no reports of equilibrium coexistence \citep{filion2010crystal, wohler2022hard}. Even large, long simulations of truly hard spheres do not report explicit phase separation \citep{pieprzyk2019thermodynamic}. Experimental estimates imply that, for $\mathcal{O}(10^6)$ particles, generating spontaneous, durable phase separation would take $\sim 3.17\times 10^8$ years \citep{tenwolde1996numerical}, i.e., effectively unreachable in simulation.

To traverse the tie line in finite time with minimal perturbation, we simulate the weakest practicable departure from metastability in a very large MPRHS system. We enforce a purely entropic competition (no attractions, no gravity) and preserve the long‑range/short‑range entropy exchange by strictly imposing a single  particle size. We approach the LLN by using $N=2{,}000{,}000$ spheres. To gently perturb metastability---while connecting to prior work---we introduce widely distributed crystal seeds totaling $0.5\%$--$4\%$ crystalline fraction. Our perturbations do {\em not} aim to change the equilibrium phase behavior; they are used to slightly destabilize the long-lived metastable fluid or crystal, so that phase separation can actually occur on finite simulation time scales. This ``weak triggering'' allows us to observe and quantify how Frenkel's entropy-exchange mechanism operates in a system that otherwise behaves as an MPRHS fluid. By contrast, strong triggers \citep{auer2001prediction, auer2001suppression, auer2004numerical, filion2010crystal, filion2011simulation, hermes2011nucleation, espinosa2016seeding, fiorucci2020effect, montero2020young, montero2020interfacial, gispen2024finding, ladd1977triple, davidchack1998simulation, noya2008determination, zykova2010monte, espinosa2013fluid, tateno2019influence, sanchez2021fcc}---such as direct-coexistence protocols \citep{ladd1977triple, davidchack1998simulation, noya2008determination, zykova2010monte, espinosa2013fluid, tateno2019influence, sanchez2021fcc} that insert a large crystal slab into the fluid, effectively placing the system right at the coexistence line---bypass this mechanism entirely: they are valuable for other questions but cannot interrogate the sought-after entropy-exchange mechanism. By distributing many tiny crystallites throughout the volume, we better mimic the natural competition between local mobility and long‑range entropy driven by Brownian motion.

This controlled, weak perturbation provides a finite‑time route along the coexistence tie line. It also enables us to examine a second, more fundamental factor: particle hardness. 
SM-R3C0)The intellectual merit of the work is that, by controlling and minimizing these perturbations, we can interrogate Frenkel’s proposed entropy-exchange mechanism in the idealized MPRHS limit. Our focus is dynamic and mechanistic: how phase separation proceeds once metastability is very gently broken, not whether the thermodynamic transition exists.

\section{Methods}
\label{sec:methods}

\subsection{Model system}
\label{subsec:model}

The computational model system studied here comprises 2{,}000{,}000 neutrally buoyant colloidal hard spheres of monodisperse radius $a$ suspended in a Newtonian solvent of density $\rho$ and viscosity $\eta$. Particle interactions and Brownian motion disturb the surrounding fluid with motion governed by the Stokes equations, owing to the vanishingly small Reynolds and Stokes numbers associated with the small size of colloids, $Re = \rho U a / \eta \ll 1$ and $St = (\rho_p/\rho)\,Re \ll 1$, where $U$ is a characteristic particle velocity set by Brownian diffusion. The phase behavior of purely repulsive hard colloids is controlled solely by the colloid volume fraction, $\phi = 4\pi a^3 n / 3$, where $n$ is the number of colloids per unit total volume. Each particle experiences hydrodynamic drag and Brownian forces as described below. Many-body hydrodynamic interactions are neglected. The systems studied lie in the volume-fraction range $0.49 \le \phi_{\textrm{target}} \le 0.55$, spanning the entire theoretical coexistence region.

\noindent \textbf{Interaction potential and nearly hard-sphere limit.} Our goal is to model nearly hard-sphere colloids under Brownian dynamics, rather than mathematically ideal hard spheres. To this end, we use a short-ranged Morse pair potential with parameters chosen so that the reduced second virial coefficient differs from the hard-sphere value by at most $1\%$. In this sense, the Morse interaction is a quantitatively accurate hard-sphere surrogate that also provides a clean control parameter for particle softness (through the potential depth and range). Throughout, we deliberately exploit this tunability to study how small, explicitly quantified deviations from the MPRHS limit affect phase separation and Frenkel’s entropy-exchange mechanism.

To represent the hard-sphere condition in simulation, entropic exclusion was modeled using a purely repulsive pair potential \(V(r)\), where \(r\) is the center-to-center distance between particles. To avoid a singular contact condition, we employed a short-range Morse potential with strong repulsion, truncated at contact:
\begin{equation}
V(r) =
\begin{cases}
    -V_0 \!\left( 2\,\mathrm{e}^{-\kappa [r-(a_i+a_j)]} - \mathrm{e}^{-2\kappa [r-(a_i+a_j)]} \right), & r \le a_i+a_j,\\[4pt]
    0, & r > a_i+a_j.
\end{cases}
\label{eqn:morse}
\end{equation}

Equation~\ref{eqn:morse} describes a nearly hard-sphere interaction between particles \(i\) and \(j\). The potential hardness is controlled by the parameters \(V_0\) and \(\kappa^{-1}\), with larger values corresponding to steeper repulsion. The baseline parameters \(V_0 = 6kT\) and \(\kappa = 30/a\), together with the exponential form of the Morse potential, have been widely used to approximate hard-sphere behavior in colloidal simulations of diffusion, flow, and gelation \citep{zlr-14,jlz-18,johnson2021phase,ryu2022modeling,lrz-16,pz-18,johnson2019influence,aponte2016simulation,aponte2018equilibrium,gonzalez2021impact,aponte2022confined,sunol2023confined}. The attractive part of \(V(r)\) was truncated to yield a purely repulsive potential. Under these parameters, the reduced second virial coefficient is \(B_2^* \equiv B_2/B_2^{HS} = 0.985\), where \(B_2^{HS}\) denotes the hard-sphere reference value. A value of \(B_2^*=1\) defines the ideal hard-sphere limit; thus \(B_2^* = 0.985\) corresponds to an effective particle deformation of only 1--2\%, consistent with experimental estimates for PMMA or polystyrene colloids \citep{rpw-13}. 

\begin{figure}
	\vspace{-0mm}
	\centering
	\includegraphics[width=0.91\linewidth]{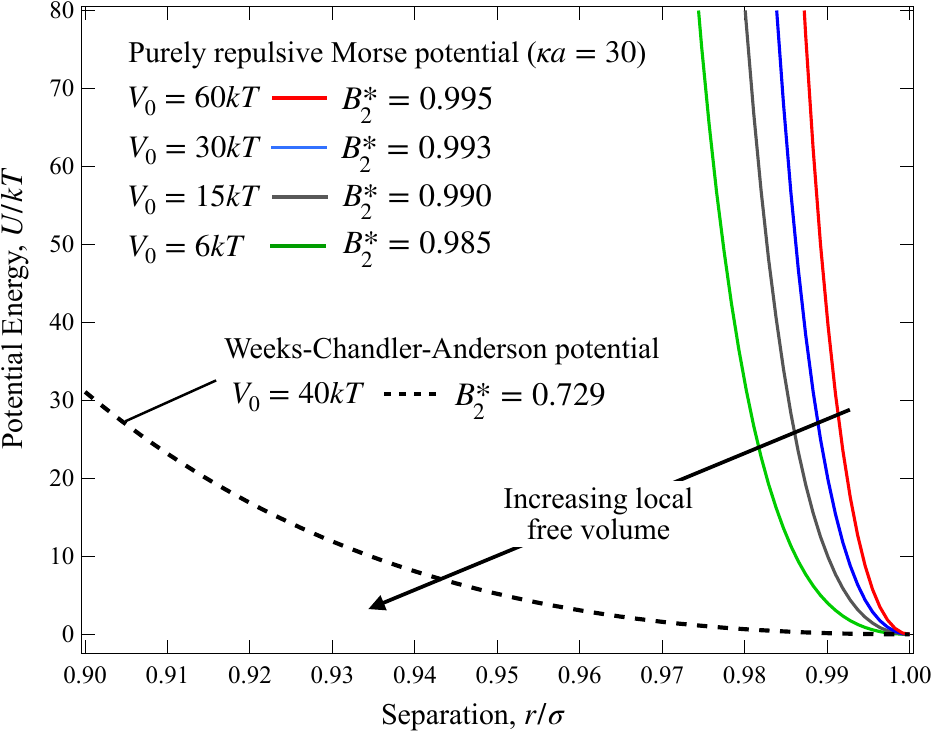}
	\caption{Comparison of potentials used to represent hard-sphere colloids in simulations, plotted as a function of particle center-to-center distance, where values smaller than unity indicate `overlap'.  The  purely repulsive Morse potential with $\kappa a=30$ (solid lines) is shown for varying hardness values as indicated in the legend. A commonly-used Weeks-Chandler-Anderson potential is also shown (black dashed line). Truly hard-sphere interaction is a Heaviside function at unity.}
	\label{fig:fig_potential}
\end{figure}

In this study we systematically explored perturbations of this nominally hard-sphere condition. The potential depth $V_0$ was varied from $6kT$ to $15kT$, $30kT$, and $60kT$, corresponding to increased hardness with $B_2^* = 0.990$, $0.993$, and $0.995$, respectively (Fig.~\ref{fig:fig_potential}). These values closely match experimental estimates for poly(12-hydroxystearic acid)–stabilized PMMA particles, which exhibit $0.969 \le B_2^* \le 0.999$~\citep{bryant2002hard}. We intentionally employ an “extremely hard” but finitely soft Morse potential as a controlled approximation to hard spheres, not as an exact hard-sphere model. In this work we deliberately vary particle hardness as a tunable perturbation away from the ideal MPRHS limit in order to obtain phase separation on accessible time scales and to probe the entropy-exchange mechanism. For the parameters used here, the Morse system remains quantitatively close to hard-sphere thermodynamics while primarily modifying the kinetics.

For comparison, many colloidal simulations employ the Weeks-Chandler-Andersen (WCA) potential as a nominally hard-sphere model (see Appendix~\ref{sec:app_hardness}). As shown in Fig.~\ref{fig:fig_potential}, even our softest Morse potential (\(B_2^* = 0.985\), \(V_0 = 6kT\)) produces substantially steeper repulsion than the WCA potential. The WCA form with \(V_0 = 40kT\) used in several prior studies \citep{filion2011simulation,fiorucci2020effect,gispen2024finding,tateno2019influence} yields \(B_2^* = 0.729\), permitting roughly 30\% effective particle overlap, far softer than the 2\% deformation typical in experiments \citep{rpw-13}. Such soft interactions shift the phase envelope to higher volume fractions, a well-established effect in the literature \citep{hoover1971thermodynamic,Robbins1988,Meijer1991,piazza1993equilibrium,Lowen1993,Lowen1993b,nemeth1995solid,senff1999temperature,likos2001effective,castelletto2002liquid,Laurati2005,archer2005density,likos2006soft,mladek2007phase,mladek2007clustering,Vlassopoulos2014,gupta2015dynamic,pelaez2015impact,Zakhari2017,erigi2023phase}. Consequently, these WCA-based systems typically require rescaling of the freezing point, which in turn misaligns the predicted melting point \citep{poon2012measuring}.

\subsection{Dynamic simulation model and algorithm}
\label{subsec:sim}
\noindent \textbf{Brownian dynamics and hydrodynamics.} All simulations are performed using overdamped Brownian dynamics with hydrodynamic interactions, appropriate for colloidal particles suspended in a solvent. In this framework, solvent-mediated thermal fluctuations set both short- and long-time self-diffusion and drive the configurational rearrangements and local vibrational motion that govern phase behavior. Our aim is therefore not to reproduce the event-driven dynamics of atomic hard spheres, but to capture the kinetics of nearly hard-sphere colloids under a realistic Brownian description while systematically varying particle hardness. 

By contrast, Monte Carlo (MC) schemes, including event-chain MC, are designed for algorithmically driven equilibration: trial moves (which may be large, collective, or chain-like) are constructed to accelerate exploration of configuration space and convergence to the correct equilibrium ensemble, rather than to mimic Brownian or Newtonian trajectories. Such methods are ideal for determining equations of state and coexistence boundaries, and we take the hard-sphere thermodynamics established by these approaches as given. They are not, however, intended to resolve the actual time scales and pathways by which a Brownian suspension reaches fluid–crystal coexistence from a homogeneous initial state, which is the kinetic question we address here.

We conduct Brownian Dynamics simulations utilizing the LAMMPS molecular dynamics package \citep{thompson2022lammps} which has a parallelization scheme optimized to handle large particle systems. We distributed 2,000,000 particles, all of size $a$, throughout the simulation cell. To efficiently initialize the system with high volume fraction, we placed all particles on a periodic lattice, then allowed its configuration to relax via Brownian motion throughout simulation. The simulation cell is replicated into an infinite domain.

LAMMPS' implicit solvent package solves the Langevin equation for each particle at each time step throughout simulation:
\begin{equation}
	\textbf{m} \cdot \frac{d\textbf{\textit{U}}}{dt} = \textbf{\textit{F}}^H+\textbf{\textit{F}}^B+\textbf{\textit{F}}^P.
	\label{eqn:langevin}
\end{equation}
Here, $\textbf{F}^H$, $\textbf{F}^B$, and $\textbf{F}^P$ are the  Stokes drag, the stochastic Brownian force, and the interparticle forces, respectively. Although many-body hydrodynamic interactions play a role in suspension mechanics even up to volume fractions as high as $55\%$ \citep{zss-15}, in cases where  repulsion keeps particles' no-slip surfaces separated by at least twenty percent of their size, these interactions become weak and can be neglected to good approximation \citep{bbv-02,kb-06,ksb-06,sb-07}. Making this freely-draining  approximation, the hydrodynamic force on each particle is determined by  Stokes' drag law:
\begin{equation}
	\textbf{\textit{F}}_i^H = -6\pi\eta a_i \left[ \textbf{\textit{U}}_i-\textbf{\textit{u}}^{\infty}(\textbf{\textit{X}}_i) \right].
	\label{eqn:hydro}
\end{equation}
Here, $\textbf{\textit{U}}_i-\textbf{\textit{u}}^{\infty}(\textbf{\textit{X}}_i)$ represents the particle velocity $\textbf{\textit{U}}_i$ relative to the fluid velocity $\textbf{\textit{u}}^{\infty}(\textbf{\textit{X}}_i)$. The Brownian force obeys Gaussian statistics \citep{bbk-84}:
\begin{equation}
	\overline{\textbf{\textit{F}}_i^B} = 0\textrm{, }\overline{\textbf{\textit{F}}_i^B(0)\textbf{\textit{F}}_i^B(t)} = 2kT(6\pi\eta a_i) \textbf{\textit{I}}\delta(t),
	\label{eqn:Brownian}
\end{equation}
where the overbars indicate averaging over a time period larger than the solvent timescale and $\textbf{I}$ is the identity tensor. The Dirac delta distribution $\delta(t)$ indicates that the Brownian impacts are instantaneously correlated. The interparticle force is defined as the negative gradient of the interparticle potential $V(r)$, and because the Morse potential is spherically symmetric, we incorporate its derivative in the spherical coordinate system:
\begin{equation}
	\textbf{\textit{F}}_i^P = - \sum_{j} \frac{\partial V(r_{ij})}{\partial r_{ij}} \hat{\textbf{\textit{r}}}_{ij}.
	\label{eqn:interparticle}
\end{equation}
Here, $\hat{\textbf{\textit{r}}}_{ij}=\textbf{r}_{ij}/r_{ij}$, where $\textbf{r}_{ij}=\textbf{X}_{i}-\textbf{X}_{j}$ is the separation vector from the center of particle $i$ to the center of particle $j$, and $r_{ij}=|\textbf{r}_{ij}|$. The summation is taken over all interacting pairs involving particle $i$. In LAMMPS, particle velocities and positions are advanced in time numerically using velocity Verlet integration \citep{at-87}. To model colloidal physics, the Reynolds number and the Stokes number must be small; in LAMMPS, this requires thoughtful selection of the integration time step, which we set at $\Delta t=10^{-5}a^2/D$, where $a^2/D$ is the diffusive time required for a single particle of size $a$ diffusing its size in pure solvent with diffusion coefficient $D=kT/6\pi\eta a$. The small time step permits only very small particle overlaps, which are resolved via a standard Heyes-Melrose algorithm \citep{hm-93}. This overlap resolution represents an entropic encounter that contributes appropriately to the osmotic pressure \citep{zb-12,zlr-14}.

We explored phase behavior within the theoretical coexistence region by preparing 13 samples at target volume fractions spanning \(0.49 \le \phi \le 0.55\), then monitoring the evolution of crystal fraction and osmotic pressure over time. Phase transitions were induced using both freezing and melting protocols to expose possible path dependence, where we expected metastable systems that retained their initial fluid structure well above the freezing point and crystalline structure well below the melting point. Similar to the asymmetric approach signatures observed in glasses \citep{kovacs-64,dwmnlpc-11,dpm-14,pm-16,mckenna2020looking}, non-linear kinetics, where particle mobility changes as \(\phi\) varies, introduce hysteresis in rate-dependent processes, potentially assisting the system in escaping metastability and reaching equilibrium coexistence. If the final, phase-separated state (same final crystal fraction) is identical for both freezing and melting protocols, the resulting state can be identified as the equilibrium coexistence condition.

\noindent\textbf{\small Preparation.}
Freezing and melting protocols were designed to prepare samples on or near the metastable fluid and crystal lines with well-controlled initial crystal fractions. Because a pristine MPRHS system would require infinite time to phase separate \citep{frenkel-93}, we deliberately introduced small, spatially distributed crystal seeds, enabling observation of phase behavior on finite timescales. The goal was to mimic the natural emergence of nucleites from thermal fluctuations while avoiding strong seeding effects, such as the use of a single crystalline substrate, that can artificially force or suppress metastability \citep{auer2001prediction,auer2001suppression,auer2004numerical,filion2010crystal,filion2011simulation,hermes2011nucleation,espinosa2016seeding,fiorucci2020effect,montero2020young,montero2020interfacial,gispen2024finding,ladd1977triple,davidchack1998simulation,noya2008determination,zykova2010monte,espinosa2013fluid,tateno2019influence,sanchez2021fcc}. Our approach thus introduces the smallest practicable perturbation to spontaneous phase separation.

\noindent\textbf{\small Freezing protocol.}
The system was initialized as a face-centered-cubic (FCC) lattice at a volume fraction of \(\phi_{0,fr} = 0.45\), from which Brownian motion immediately began to relax the structure. The simulation box was then uniformly contracted to increase the packing fraction from 0.45 to 0.56 at controlled rates. This densification was performed either slowly or rapidly relative to the Brownian relaxation time, producing samples with fewer or more distributed crystal seeds, respectively. Two freezing rates were applied:
\begin{equation*}
d\phi/dt \vert_{\text{freezing}} = 
\begin{cases}
0.025\,D/a^2 \quad &(\text{slow, total time } 4\,a^2/D),\\
0.25\,D/a^2 \quad &(\text{fast, total time } 0.4\,a^2/D).
\end{cases}
\end{equation*}
At higher densification rates, Brownian relaxation becomes less effective, retaining more of the initial FCC order. The result was two distinct sets of 13 samples: one near the metastable fluid line (slow freezing, \(\zeta_0 = 2\text{--}4\%\) crystal fraction) and one near the metastable crystal line (fast freezing, \(\zeta_0 = 84\text{--}99\%\)).

\noindent\textbf{\small Melting protocol.}
For melting tests, the system was first compressed to \(\phi = 0.56\) and allowed to relax for \(4\,a^2/D\), yielding an initial crystal fraction of 2--4\%. The simulation box was then expanded to decrease \(\phi\) at two controlled rates:
\begin{equation*}
d\phi/dt \vert_{\text{melting}}  = 
\begin{cases}
-0.025\,D/a^2 \quad &(\text{slow}),\\
-0.1\,D/a^2 \quad &(\text{fast}).
\end{cases}
\end{equation*}
As in freezing, varying the expansion rate tuned the number of remnant crystal seeds. Two corresponding sets of 13 samples were thus prepared along the metastable fluid line, each containing tiny, distributed crystalline seeds.

\noindent\textbf{\small Quasi-equilibrium protocol.}
A third, quasi-equilibrium series was prepared by melting while allowing the system to relax at each step by \(2{,}000\,a^2/D\) after every incremental decrease \(\Delta \phi = -0.01\). This approach approximates an equilibrium path through the coexistence region.

\noindent\textbf{\small Equilibration and monitoring.}
Upon reaching each target volume fraction (the frozen or melted configuration), the system was held at constant \(\phi\) and evolved via Brownian dynamics under isothermal conditions (Eq.~\ref{eqn:langevin}). Osmotic pressure and crystal fraction were monitored until both reached steady values. Equilibrium plateaus were typically achieved within \(t/(a^2/D) = 2{,}000\text{--}8{,}000\); the minimum monitoring time was therefore set to \(2{,}000\,a^2/D\).

\noindent\textbf{\small Hardness variation.}
All protocols described above were repeated for four values of particle hardness to quantify how small variations in accessible free volume mediate the exchange between long-range (configurational) and short-range (vibrational) entropy. \\

\begin{figure*}[t!]
	\vspace{-0mm}
	\centering
	\includegraphics[width=0.95\linewidth]{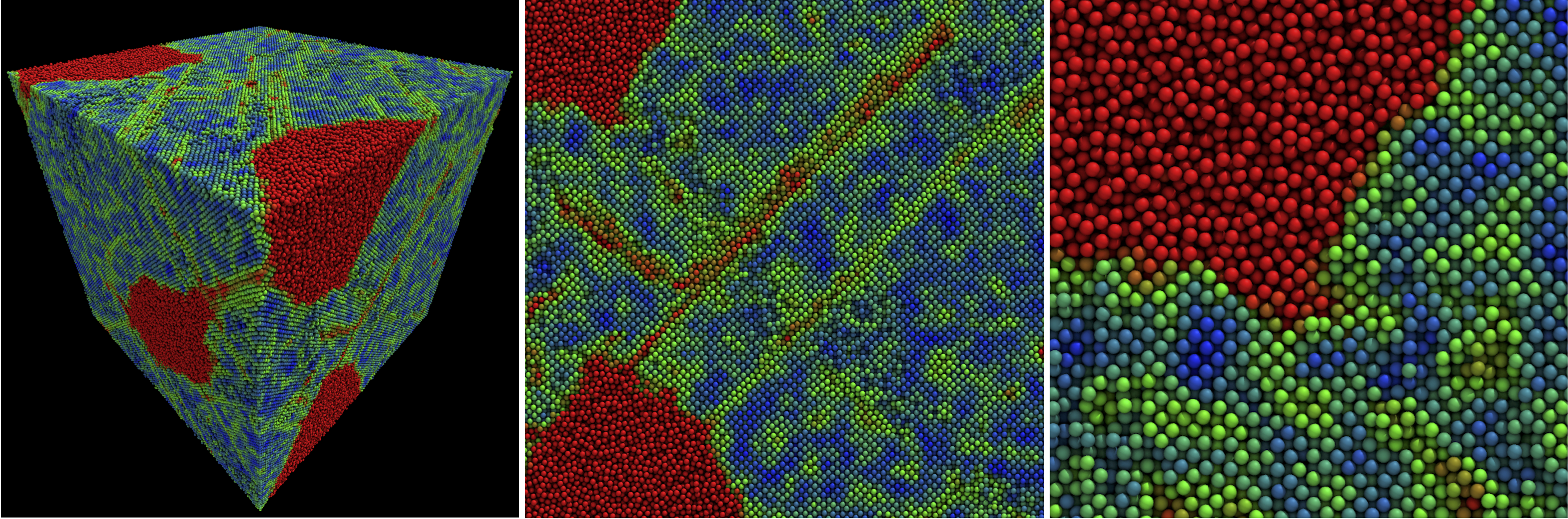}
	\caption{{Snapshots of our Brownian dynamics simulations of the phase behavior of solvent-suspended colloids. Far left: simulation cell of 2,000,000 colloids, replicated periodically into an infinite domain in LAMMPS \citep{thompson2022lammps}. Second and third images: same system at 2x and 5x magnification. Colors correspond to local order, ranging from red for structureless to deep blue for perfect crystal structure. Figure from Wang {\em et al.}\citep{wangInreviewElusive}, with permission.}}\label{fig:fig_snapshots}
	\vspace{-0mm}
\end{figure*}

\subsection{Structure and osmotic pressure measurement}
\label{subsec:measurements}
We track the positions, velocities, and particle-phase stress throughout the freeze or melt processes. We measure the radial distribution function, then use it to quantify the extent of crystallization in our calculation of the per-particle bond-orientational order parameters $\bar{q}_6$ and $\bar{q}_4$, from which we calculate the crystal fraction at any time during the freeze and melt process. Using this data we plot the crystal fraction as a function of volume fraction to deduce regions of a colloidal phase diagram.

The average local-order parameter is defined, for a particle $i$ with a number of neighboring particle $N_b$, as \citep{Lechner2008, Steinhardt1983}
\begin{equation}
  \bar q_l(i)=\sqrt{\frac{4\pi}{2l+1}\sum^l_{m=-l}|\bar q_{lm}(i)|^2},
\end{equation}
where
\begin{equation} \label{Eq:OP-2}
  \bar q_{lm}(i)=\frac{q_{lm}(i)+\sum_{k=1}^{N_b}q_{lm}(k)}{N_b+1},
\end{equation}
and
\begin{equation}
  q_{lm}(i)=\frac{\sum_{j=1}^{N_b}Y_{lm}(\mathbf{r}_{ij})}{N_b}.
\end{equation}
Here, $q_{lm}$ is a complex number depending on all spherical harmonics $Y_{lm}$ of order $l$ and where integers $m \in \{-l,\ldots,l\}$, for a pair of particles with center-to-center vector separation $\mathbf{r}_{ij}$. 

In Eq.~(\ref{Eq:OP-2}), $q_{lm}$ is averaged over both particle $i$ and its neighbors $N_b$, enhancing the ability to distinguish between different crystal structures~\citep{Lechner2008}. Particles are considered neighbors if their separation is less than the distance corresponding to the first minimum of the radial distribution function. The absolute value of the crystal fraction is, in principle, cutoff-dependent, because changing the cutoff alters which neighbors contribute to $q_6$ and thus which particles are tagged as ``solid-like''. If the cutoff is too small, true first-shell neighbors are missed and genuinely crystalline particles can fall below the solid-like threshold; if it is too large, second-shell or fluid-like neighbors contaminate the bond-orientational signal and can either inflate or deflate the apparent crystal fraction. In our analysis, we therefore define neighbors using the first minimum of $g(r)$ at each volume fraction, which isolates the first coordination shell and minimizes contamination from more disordered neighbors. Within a narrow range around this first minimum, our qualitative conclusions about the magnitude and trends of the crystal fraction are robust.

The spherical harmonics of orders $l=4$ and $l=6$ are used in the present study to identify structures with four-fold symmetry, such as body-centered cubic (BCC), and six-fold symmetry, for hexagonal close packed (HCP) and face-centered cubic (FCC), respectively. Particles are classified as crystalline if $\bar q_6\ge0.29$ and further categorized as BCC for $\bar q_4\le0.05$, HCP for $0.05<\bar q_4\le0.1$, and FCC for $\bar q_4 >0.1$ \citep{Lechner2008, Kratzer2015}. Based on the average local-order parameter, the structure can be further quantified in terms of fractions of BCC, HCP and FCC crystals as well as the fluid (amorphous) phase.

In the Results section, we will report the particle-phase osmotic pressure in connection with phase behavior. Osmotic pressure is defined as the negative of one third of the trace of the stress. The particle-phase stress $\mathbf\Sigma^P$ in a freely-draining suspension arises from the presence of the particles --- the ideal osmotic pressure --- as $nkT\textbf{I}$, and the interparticle elastic stress $\textbf{\textit{r}} \textbf{\textit{F}}^P$ due to interactions:
\begin{equation}
	\langle \mathbf{\Sigma} \rangle = -nkT \textbf{\textit{I}} - n \left<\textbf{\textit{r}} \textbf{\textit{F}}^P\right>.
	\label{eqn:stress}
\end{equation}
Here, \textbf{I} is the identity tensor, $\textbf{\textit{r}}$ is the center-to-center distance between an interacting pair, and the angle brackets indicate an average over all particles.   This particle phase stress plus the solvent stress give the total suspension stress $\langle\boldsymbol{\sigma}\rangle$\citep{batchelor1977effect,fb-00,brady1993brownian, zb-12}. 

The osmotic pressure in a suspension also includes both the contribution due to solvent thermodynamic pressure and that arising from the presence of the particles, their diffusion, and interactions between the particles --- the particle-phase osmotic pressure $\langle\Pi^P\rangle$:
\begin{equation}
	\langle\Pi^P\rangle = -\frac{1}{3} \textbf{I}: \langle \mathbf{\Sigma}^P \rangle.
	\label{eqn:op}
\end{equation}

\subsection{Free energy, hard spheres, and osmotic pressure}
\label{subsec:free_energy}

At fixed temperature, spontaneous phase separation in colloidal dispersions is governed by minimization of the Helmholtz free energy,
\begin{equation*}
A = U - TS,
\end{equation*}
where \(U\) is the internal energy, \(T\) the absolute temperature, and \(S\) the entropy \citep{balescu1975equilibrium,russel1991colloidal}. Coexistence between a fluid and a crystal is obtained from the standard equalities of intensive variables,
\begin{equation*}
\mu_{\mathrm{fluid}}(T,\phi_{\mathrm{fluid}})=\mu_{\mathrm{solid}}(T,\phi_{\mathrm{solid}}), 
\end{equation*}
\begin{equation*}
\Pi_{\mathrm{fluid}}(T,\phi_{\mathrm{fluid}})=\Pi_{\mathrm{solid}}(T,\phi_{\mathrm{solid}}),
\end{equation*}
which define the tie line connecting the fluid and solid branches of the phase envelope. {The thermodynamic connection among free energy \(A\), osmotic pressure \(\Pi\), and chemical potential \(\mu\) follows from extensivity: \(A=\mu N-\Pi V\)}, showing that the osmotic pressure $\Pi$ is related thermodynamically to the {chemical potential} $\mu$ as $\mu/kT = A/NkT+\Pi/nkT$, representing the increase in pressure and the energy per unit volume required to add another particle to a system of fixed size, respectively. Here, $N$ is the total number of particles in the volume $V$, $n\equiv N/V$ is the number density, and $kT$ is the thermal energy.

\begin{figure*}[ht]
	\vspace{0mm}
	\centering
	\includegraphics[width=0.9\textwidth]{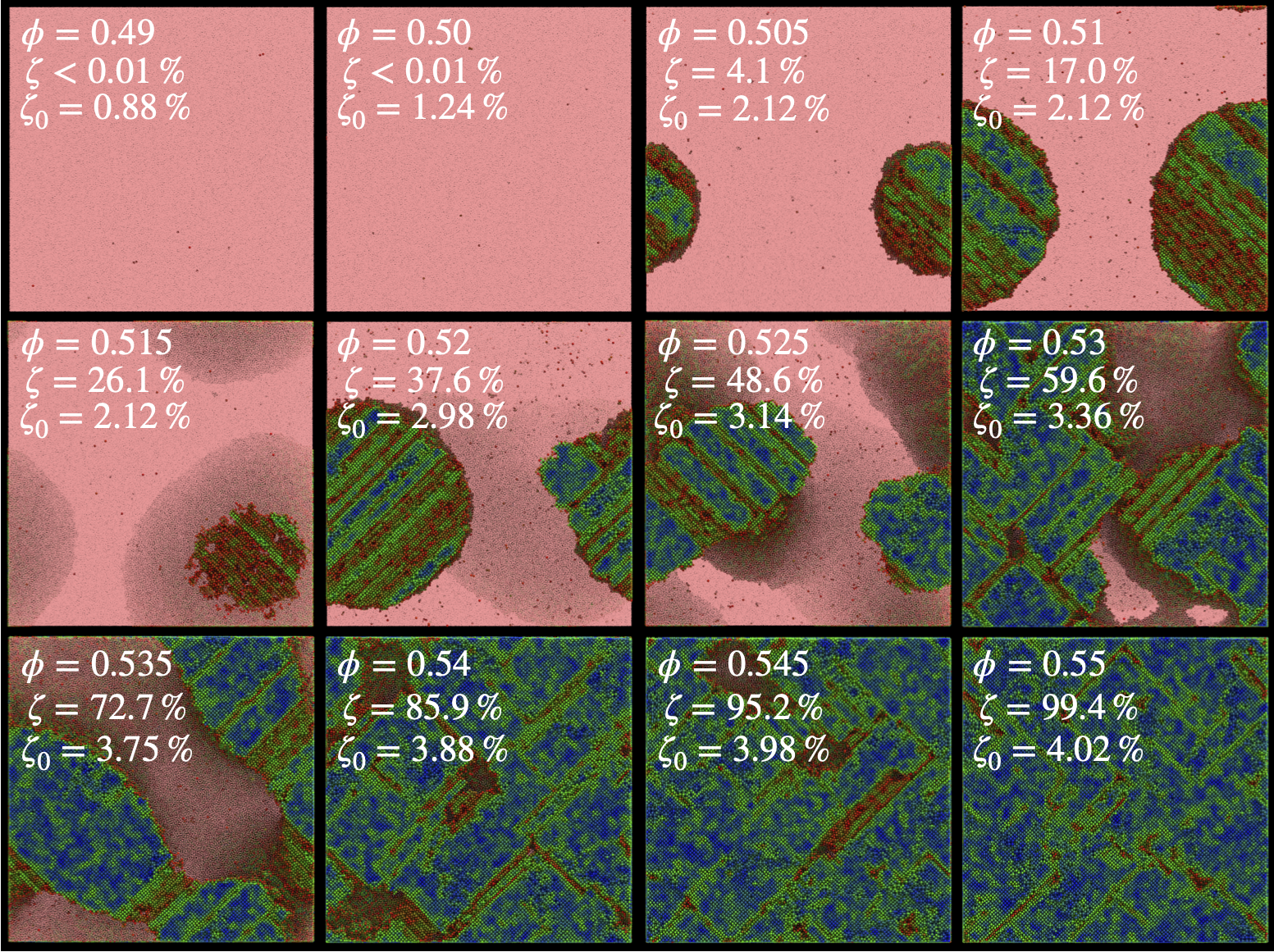}
	\caption{Simulation images from present study showing particle arrangements for a range of volume fraction $\phi$ and crystal fraction $\zeta$. Particles are colored according to $6^{th}$ order average local-order parameter $\bar{q}_6$. Particles surrounded by amorphous structure ($\bar{q}_6<0.29$) are colored pink and made translucent for visibility. Red particles are surrounded by marginally crystalline structure ($\bar{q}_6\approx0.3$); green particles are surrounded by substantially crystalline structure ($\bar{q}_6\approx0.4$); and blue particles ($\bar{q}_6\geq0.5$) are surrounded by very crystalline structure. Particle hardness set as $V_0=6kT$ and $\kappa a=30$. All images from samples initially close to the theoretical metastable fluid line (all using the slow melting protocol, except for $\phi=0.505$ and $\phi=0.51$ that used the quasi-equilibrium melting protocol).}\label{fig:fig_image_panel}
	\vspace{0mm}
\end{figure*}

For purely repulsive hard spheres, the pair potential satisfies \(V(r)=\infty\) for \(r<2a\) and \(V(r)=0\) for \(r\ge 2a\). Configuration space is therefore constrained solely by excluded volume, and the free energy is entropic in origin. Consistently, the reduced second virial coefficient \(B_2^\ast\) approaches unity in the hard-sphere limit. In our simulations we approximate the hard core with a steep, short-ranged, \emph{repulsive} Morse form. Tuning its amplitude controls an effective ``hardness''; as the amplitude increases, \(B_2^\ast \to 1\), approaching the hard-sphere benchmark.

Finite hardness (softness) slightly relaxes the exclusion constraint: small deformations at contact increase locally accessible volume and open additional short-range (vibrational) configurations. This raises vibrational entropy and facilitates rearrangements needed to accommodate an additional particle. Conversely, as particles become harder, the set of available configurations at fixed \(\phi\) shrinks; adding a particle then requires rarer, cooperative rearrangements that create a full particle-sized cavity, which elevates the osmotic pressure. At equal volume fraction we therefore expect
\begin{equation*}
\Pi_{\text{soft}}(\phi) \;<\; \Pi_{\text{HS}}(\phi),
\end{equation*}
with corresponding shifts of the fluid and crystal branches of the metastable phase diagram relative to the atomic hard-sphere reference.

These considerations yield two testable expectations for our nearly hard, monodisperse systems: (i) minimal, distributed crystalline seeds plus slight softness should enable explicit, long‑lived fluid-crystal coexistence along the tie line in finite time; and (ii) even without seeding, tiny softness should produce spontaneous coexistence within a narrower window of \(\phi\), consistent with  a (metastable) coexistence region in the strict hard-sphere limit where the (stable) phase separated state is dynamically inaccessible. We now present results that quantify these effects on crystalline structure,  inferred coexistence boundaries and osmotic pressure.

\section{Results and discussion}\label{sec:results}
\begin{figure*}[t]
	\vspace{0mm}
	\centering
	\includegraphics[width=0.9\textwidth]{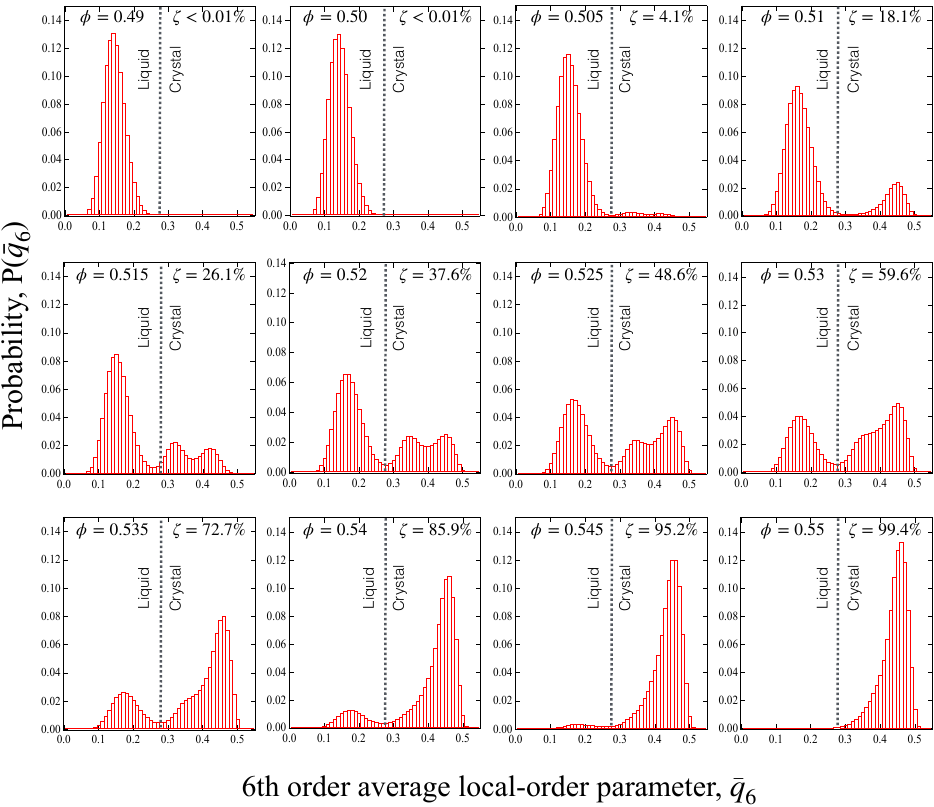}
	\caption{{Extent of crystal and fluid-like structure at 12 volume fractions as shown. Total crystal fraction $\zeta$ shown in each plot. The probability P($\bar{q}_6$) is plotted as a function of the {$6^{th}$ order average local-order parameter} $\bar{q}_6$, calculated for each of the 2,000,000 particles. Measurement taken at $2,000a^2/D$ after achieving target volume fraction. Dotted vertical line marks the boundary between fluid-like structure ($\bar{q}_6 < 0.29$) and crystalline structure ($\bar{q}_6 \ge 0.29$). Particle hardness $V_0=6kT$, $\kappa a=30$ ($B_2^*=0.985$), cf Figure \ref{fig:fig_potential}. }}
	\label{fig:fig_q6_hist}
	\vspace{0mm}
\end{figure*}
We report measurements of crystal fraction, phase envelope, and osmotic pressure for sets of thirteen samples prepared at target volume fractions spanning the MPRHS phase envelope. As detailed in the \S\ref{sec:intro} and \S\ref{sec:methods}, and consistent with prior literature, spontaneous equilibrium phase separation is not expected in pristine systems of monodisperse, purely repulsive hard spheres \citep{frenkel-93}. The metastability of such systems---whether along the fluid or solid line---would require astronomical timescales to relax, even for extremely large systems \citep{tenwolde1996numerical}.

Instead, we probe this metastable regime by preparing samples very near a metastable line, each containing distinct, spatially distributed crystal seeds. Samples prepared near the metastable \emph{fluid} line were generated by controlled melting or slow freezing and contained 2--4\% distributed crystal nuclei. Samples prepared near the metastable \emph{solid} line were generated by rapid freezing and contained 84--99\% crystalline material. The precise initial crystal fraction, \(\zeta_0\), is reported alongside the data for each final state.

To interrogate Frenkel’s proposed long-range/short-range entropy-exchange mechanism, we designed simulations that isolate both competing contributions. First, the system size---2,000,000 particles---provides substantial configurational (long-range) entropy. Second, we systematically varied particle hardness across four values in the ``very hard'' regime to perturb the available short-range vibrational entropy via minute changes in local free volume around slightly deformable particles.

The following sections present structural and thermodynamic results in sequence. Section~\ref{subsec:structure} reports structural metrics, most principally crystal fraction, from which we infer the phase envelope in Section~\ref{subsec:phase}. The resulting phase diagram, expressed as osmotic pressure versus volume fraction, is discussed in Section~\ref{subsec:op}. Section~\ref{subsec:frenkeltest} examines the effect of particle hardness as a direct test of Frenkel’s mechanistic model, and Section~\ref{subsec:app_time} presents the time evolution of osmotic pressure and crystal fraction. In each case, we emphasize the path dependence of the final system state as a function of the sample preparation route (melting versus freezing).

\subsection{Structural measurements}
\label{subsec:structure}
Using the computational framework outlined above, we simulated freezing and melting of a colloidal dispersion of very hard spheres with particle hardness parameters \(V_0=6kT\) and \(\kappa a=30\), giving the reduced second virial coefficient \(B_2^*=0.985\) (baseline hardness; see Methods). A representative simulation (replicated periodically into an infinite system) is shown in Fig.~\ref{fig:fig_snapshots}, which displays the full simulation cell and two zoomed-in views.

We began our phase-behavior analysis by preparing 13 samples at target volume fractions spanning \(\phi=0.49\) to \(\phi=0.55\) using the slow-melting protocol (see Methods). For several cases we also examined quasi-equilibrium melting and fast freezing to probe protocol sensitivity. Each protocol produced a sample that was almost entirely metastable fluid with widely distributed, tiny crystal seeds; the total seed fraction for each sample is indicated in the corresponding plot. The samples at \(\phi=0.505\) and \(\phi=0.510\) were initially prepared by slow melting and remained metastable fluids for long durations. We then repeated the preparation by first relaxing the system at \(\phi=0.515\) (seed fraction \(\zeta_0=2.12\%\)) and subsequently melting down to the target \(\phi\). This quasi-equilibrium path (see Methods) allowed those two samples to phase separate, with the path dependence illustrating the system’s metastability.
\label{subsec:phase}
\begin{figure}[h]
	\vspace{0mm}
	\centering
	\includegraphics[width=0.9\linewidth]{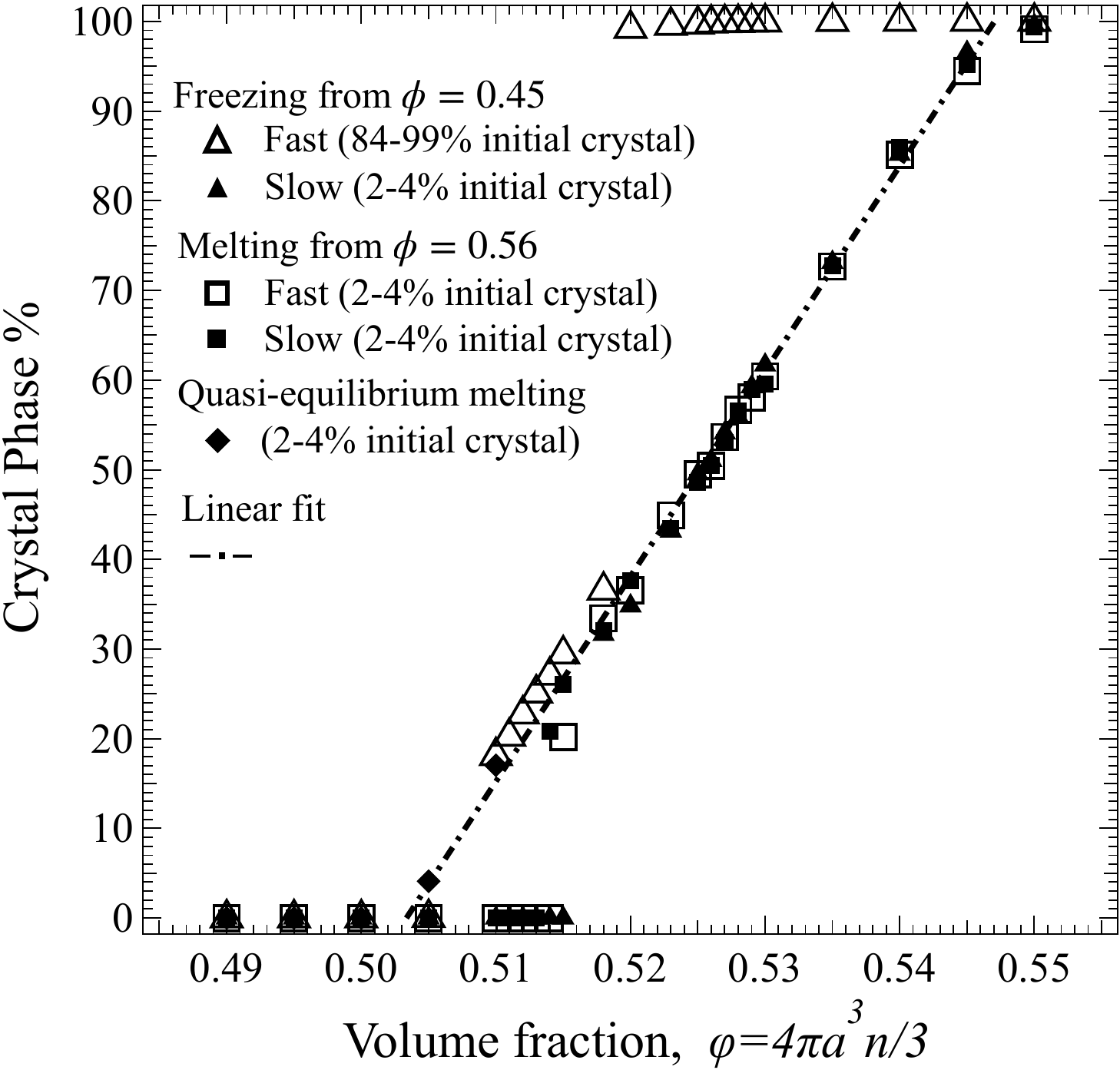}
	\caption{Crystal fraction as a function of volume fraction. Fast freezing ($\triangle$), slow freezing ($\blacktriangle$), fast melt ($\square$), slow melt ($\blacksquare$), and quasi-equilibrium melting (${\medblackdiamond}$) are shown (see Methods for rates), with the initial crystal seeding fractions also shown in the legend. A linear fit predicts freezing at $\phi=0.503$ and melting at $\phi=0.547$. Particle hardness parameters $V_0=6kT$ and $\kappa a=30$ ($B_2^*=0.985$).}\label{fig:fig_xi_6kt}
	\vspace{0mm}
\end{figure}

After each target \(\phi\) was reached, we measured the crystal fraction of every sample. A slice from the simulation box for each case is shown in Fig.~\ref{fig:fig_image_panel} together with the final crystal fraction. Particles are colored by the sixth-order averaged local bond-order parameter \(\bar{q}_6\) (see Methods). Fluid-like environments yield \(\bar{q}_6<0.29\) and are colored pink. Values \(\bar{q}_6\ge 0.29\) indicate crystalline order: marginally crystalline (\(\bar{q}_6\approx 0.30\), red), substantially crystalline (\(\bar{q}_6\approx 0.40\), green), and highly crystalline (\(\bar{q}_6\ge 0.50\), dark blue). The volume fraction \(\phi\), final crystal fraction \(\zeta\), and initial seed fraction \(\zeta_0\) are listed on each panel. For each \(\phi\), \(\bar{q}_6\) was monitored for at least \(t\ge 2{,}000\,a^2/D\), where \(D=kT/(6\pi\eta a)\) is the single-particle diffusivity. The number of colloids that attain crystalline order is statistically invariant under further Brownian evolution beyond \(1{,}000\,a^2/D\), and often stabilizes earlier (see Sec.~\ref{subsec:app_time}).

Visual inspection of Fig.~\ref{fig:fig_image_panel} shows no crystalline domains for $\phi < 0.505$. As the dispersion is effectively cooled by increasing $\phi$ into the range $0.505 \le \phi \le 0.520$, a well-defined crystalline domain emerges within a structureless fluid. Unlike direct-coexistence approaches, here the crystalline domain develops from thermal fluctuations and Brownian motion acting on the initially tiny, distributed seeds. The approximately spherical crystal is consistent with classical nucleation theory, wherein supercritical nuclei grow once a critical size is exceeded~\citep{CNT1926,debenedetti-96,Abraham2012}. With further densification to $0.525 \le \phi \le 0.545$, crystalline domains become dominant. At $\phi = 0.55$, the system is fully crystalline. Misaligned crystalline domains separated by grain boundaries are evident, likely reflecting the formation and subsequent impingement of multiple nuclei. Standard equilibrium signatures such as transitions between spherical, cylindrical, and slab-like interfacial morphologies are not observed in our simulations, in part because nucleation barriers and finite-time sampling limit access to fully equilibrated interfaces. Instead, we assess proximity to equilibrium using bulk observables: for our hardest particles, the long-time crystal fraction (\S~\ref{subsec:phase}) and pressure (\S~\ref{subsec:op}) closely match theoretical coexistence predictions for hard spheres. This agreement suggests that, despite the persistence of grain boundaries and the absence of morphology fluctuations, the system is sampling states very near the equilibrium coexistence conditions relevant to our study.

The structural composition, quantified by the distribution of \(\bar{q}_6\) across all particles, is presented in Fig.~\ref{fig:fig_q6_hist}. The dotted line at \(\bar{q}_6=0.29\) separates fluid-like (\(\bar{q}_6<0.29\)) from crystalline (\(\bar{q}_6\ge 0.29\)) environments. For \(\phi=0.49\) and \(\phi=0.50\), a single peak lies to the left of the threshold, indicating fully fluid structure; the crystal fraction \(\zeta\) (upper right of each panel) is less than \(0.01\%\). At \(\phi=0.505\), a small crystalline fraction \(\zeta=4.1\%\) appears, signaling the onset of coexistence. As \(\phi\) increases beyond \(0.505\), a second peak emerges near \(\bar{q}_6\approx 0.45\) and grows in height. Two distinct peaks---one fluid, one crystal---are observed for \(0.505\le \phi \le 0.545\), with the crystalline peak increasing in prominence in tandem with the measured \(\zeta\).

A third peak is visible only within the coexistence regime. We attribute this feature to interfacial particles located at crystal-fluid boundaries. As \(\phi\) increases, this interfacial peak shifts rightward and merges into the crystalline peak, while the fluid peak diminishes and shifts right until it disappears.

\subsection{Phase envelopes}\label{subsec:phase}
The crystal fraction data extracted from Fig.~\ref{fig:fig_q6_hist} are plotted in Fig.~\ref{fig:fig_xi_6kt} as a function of volume fraction. Several data series are shown, corresponding to two freezing protocols (fast and slow) and two melting protocols (fast and slow), as well as a quasi-equilibrium melt. These different preparation routes generate samples either near the theoretical metastable fluid line or near the theoretical metastable solid line (see Methods). The legend indicates the initial crystal seeding fraction used in each case.

The resulting ``phase diagram'' reveals a clear path dependence in the final state for volume fractions within the hard-sphere coexistence region predicted by theory \citep{hr-68,pvM-86}. Samples prepared near the theoretical metastable \emph{solid} line (via fast freezing) favor phase separation only for $\phi \le 0.518$, i.e., closer to the freezing boundary. In contrast, samples prepared near the theoretical metastable \emph{fluid} line (via slow freezing or either melting protocol) favor phase separation closer to the melting envelope and well into the coexistence region, for $\phi \ge 0.514$; At lower volume fractions, the initial metastable fluid state persisted without observable phase separation. This contrasting behavior between fluid- and solid-side preparations is analyzed further in the following section on osmotic pressure.

For all cases in which phase separation occurs---that is, when the system escapes metastability---the final state becomes path independent. This is evident from the linear alignment of all coexistence data along a single lever-rule line (dashed line in Fig.~\ref{fig:fig_xi_6kt}). A linear fit yields freezing and melting points of $\phi_{\mathrm{F}} = 0.503$ and $\phi_{\mathrm{M}} = 0.547$, respectively, defining a coexistence region between these limits. Both boundaries lie slightly above the hard-sphere, liquid-state-theory predictions for perfectly hard spheres ($\phi_{\mathrm{F}} = 0.494$, $\phi_{\mathrm{M}} = 0.545$). The origin of this small shift, linked to finite particle hardness and corresponding changes in osmotic pressure, is discussed in the next section.

\subsection{Osmotic pressure}
\label{subsec:op}
\begin{figure}[t]
	\vspace{0mm}
	\centering
	\includegraphics[width=0.91\linewidth]{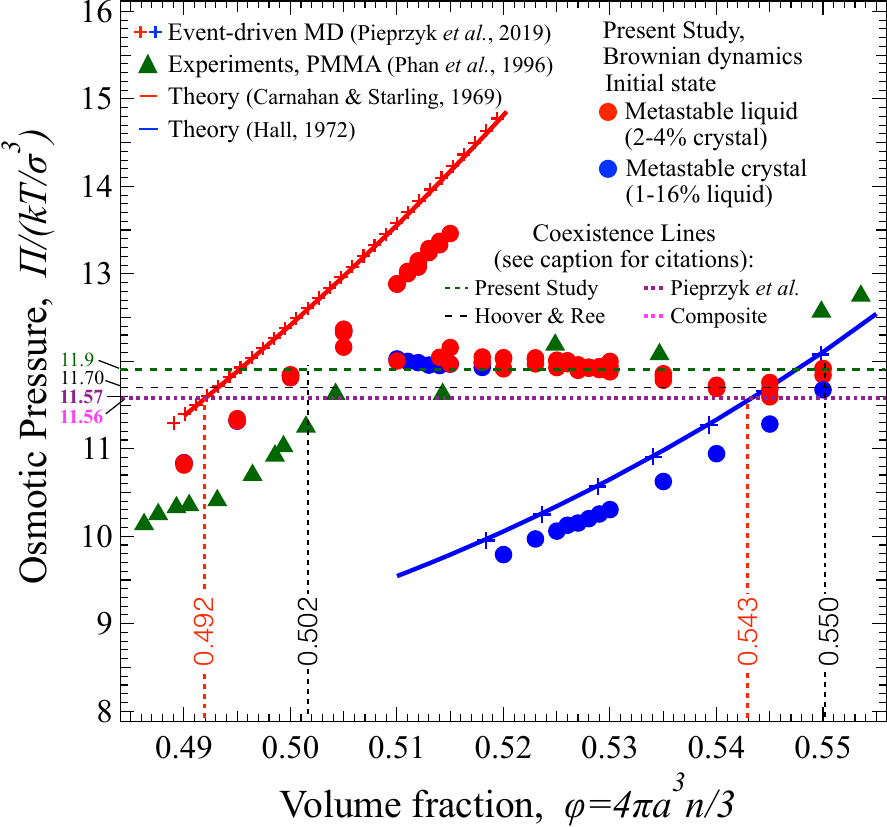}
	\caption{Osmotic pressure as a function of volume fraction in experiments, theory, and simulations for the baseline nearly hard-sphere case. Present simulations ($V_0 = 6kT$, $\kappa a = 30$, $B_2^\ast = 0.985$) were prepared near the metastable fluid and crystal lines with initial crystal fraction as shown in the legend, and yield all-fluid, all-crystal, and fluid--crystal coexistence states (red and blue circles). Theoretical predictions for the fluid branch \citep{cs-69}  and crystal branch \citep{hall1972another} for truly hard spheres are shown as solid red and blue lines and closely match the corresponding event-driven atomic simulations of \cite{pieprzyk2019thermodynamic} (red and blue crosses). Experimental data for nearly hard-sphere colloids \citep{prczcdo-96} span coexistence and pure fluid and solid phases (green triangles). Coexistence lines: the present simulation data intersect the metastable branches along the green dashed line; the Hoover--Ree coexistence pressure \citep{hr-68} is shown as a black dashed line. The magenta dotted line denotes the coexistence pressure obtained by \cite{pieprzyk2019thermodynamic}. The pink dotted line indicates the literature-average coexistence pressure compiled by \cite{royall2024colloidal}, based on multiple previous studies \citep{speedy1997pressure, davidchack1998simulation, wilding2000freezing, frenkel2002understanding, vega2007revisiting, noya2008determination, odriozola2009replica, zykova2010monte, nayhouse2011monte, fernandez2012equilibrium, ustinov2017thermodynamics, pieprzyk2019thermodynamic,moir2021tethered}. The $V_0 = 6kT$ simulation data shown here are also included in Fig.~7(b) for comparison with harder particles.}
	\label{fig:fig_pi_6kt}
	\vspace{0mm}
\end{figure}

We measured the particle-phase osmotic pressure, as described in \S\ref{sec:methods}, throughout the freezing and melting processes and during subsequent equilibration. The resulting values, averaged over all colloids, are plotted in Fig.~\ref{fig:fig_pi_6kt} for our baseline nearly hard-sphere condition (\(V_0 = 6kT\), \(B_2^* = 0.985\)). For comparison, the figure also shows results from purely repulsive hard-sphere (PRHS) experiments \citep{prczcdo-96}, hard-sphere liquid-state theory \citep{hr-68,cs-69,hall1972another}, and event-driven molecular dynamics (EDMD) simulations of atomic hard spheres \citep{pieprzyk2019thermodynamic}. Data from all initial configurations are included: red symbols represent samples prepared near the metastable fluid line, and blue symbols represent samples prepared near the metastable solid line. The combined measurements yield a fluid branch, a crystal branch, and a coexistence tie line obtained directly from coexistence mixtures.

The intersections of the fluid and crystal branches with the \emph{measured} coexistence data (filled circles), indicated by the green dashed line in Fig.~\ref{fig:fig_pi_6kt}, correspond to the condition of equal osmotic pressure between phases. From these intersections, our simulations produce a coexistence envelope bounded by \(\phi_{\mathrm{F}} = 0.502\) and \(\phi_{\mathrm{M}} = 0.550\). These phase boundaries derived from pressure equality agree closely with those inferred independently from the crystal fraction measurements (cf.\ Figs.~\ref{fig:fig_q6_hist} and \ref{fig:fig_xi_6kt}).

The fluid and crystal branches in Fig.~\ref{fig:fig_pi_6kt} show strong qualitative agreement with experiments, theory, and prior EDMD simulations. Quantitatively, our measured pressures underpredict both the fluid and crystal branches by approximately 3--4\%. This systematic offset is similar to that reported in Brownian dynamics simulations by Foss and Brady \citep{fb-00}. In contrast, our simulations slightly overpredict the coexistence pressure relative to experimental and theoretical values as well as EDMD results.

In atomic and colloidal systems alike, the osmotic pressure is the thermodynamic variable used to define phase envelopes, typically represented as pressure versus density or packing fraction. The total osmotic pressure consists of the ideal-gas contribution, $nkT$, arising from the finite-size non-interacting particles, together with contributions from entropic exclusion and higher-order interactions, all of which are built into our simulations. It was therefore unexpected that our initial simulations of very hard spheres ($V_0 = 6kT$, $B_2^* = 0.985$) yielded lower metastable-fluid and metastable-crystal pressures than predicted by hard-sphere liquid-state theory for PRHS. We do not interpret this as a discrepancy with existing theory, but as a finite-size, finite-time, and finite-softness effect: our systems are large but not in the strict thermodynamic limit, and we use a very steep but still soft Morse potential in place of an ideal hard-sphere interaction. Pushing to larger systems and harder interaction parameters is therefore expected to reduce these finite-size and finite-softness deviations and bring our simulation observables into even closer agreement with standard hard-sphere liquid-state predictions. We also note that liquid-state theoretical approaches can be and have been successfully applied to a wide range of interaction potentials, including Morse-like systems; our aim here is not to challenge those results, but to quantify how nearly hard-sphere colloids behave under our specific kinetic model. Given that our parameter set corresponds to much harder interactions than those typically employed in colloidal simulations, such as the softer Weeks–Chandler–Andersen (WCA) potentials used by the Dijkstra and Tanaka groups~\citep{filion2010crystal,filion2011simulation,tateno2019influence,fiorucci2020effect,gispen2024finding}, this underprediction implies that particle hardness in our system remains slightly perturbed relative to the pristine PRHS limit. We examine the influence of hardness more systematically in the following section.

\begin{figure*}[t]
	\vspace{0mm}
	\centering
	\includegraphics[width=0.95\linewidth]{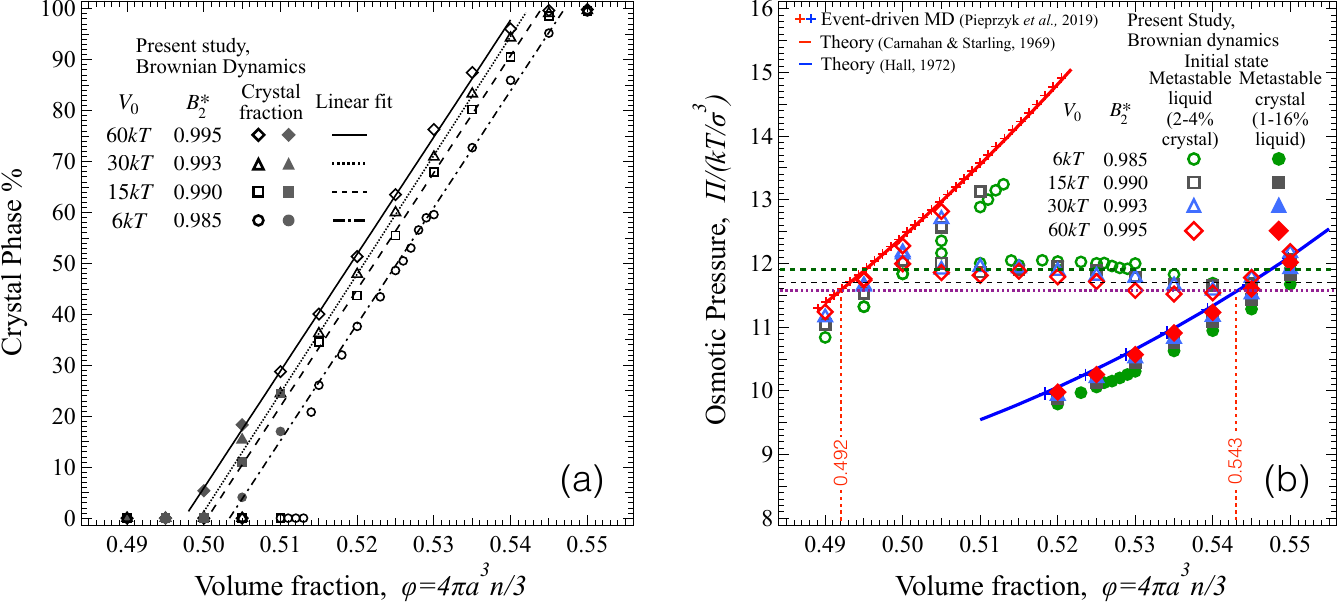}
	\caption{Impact of increasing particle hardness on phase behavior. (a) Crystal fraction as a function of volume fraction for systems with progressively harder purely repulsive Morse potentials (Eq.~\ref{eqn:morse}, with $V_0 = 6kT$ (green circles), $15kT$ (blue squares), $30kT$ (magenta triangles), and $60kT$ (orange diamonds), as indicated in the legend. All samples were prepared near the metastable fluid line (see Methods). Open symbols correspond to the melting protocol on Brownian time scales, and filled symbols to the quasi-equilibrium protocol. Solid lines are linear fits to the coexistence state for each hardness, indicating the inferred coexistence tie lines. (b) Osmotic pressure versus volume fraction for the same systems. New simulation data for increased hardness (colored symbols) are shown alongside the baseline $V_0 = 6kT$ case from Fig. \ref{fig:fig_pi_6kt}, experimental data \citep{prczcdo-96}, theoretical predictions \citep{hr-68,cs-69,hall1972another}, event-driven MD simulations, and literature-average coexistence pressures \citep{pieprzyk2019thermodynamic,royall2024colloidal} and references therein), as in Fig. \ref{fig:fig_pi_6kt}. This panel thus extends Fig. \ref{fig:fig_pi_6kt} by comparing the baseline nearly hard-sphere case to progressively harder particles.}
	\label{fig:fig_hardness}
	\vspace{0mm}
\end{figure*}

\subsection{Hardness, osmotic pressure, and the entropy-exchange mechanism}
\label{subsec:frenkeltest}

As discussed in \S\ref{subsec:free_energy}, the osmotic pressure \(\Pi\) of a colloidal suspension can be viewed as the mechanical pressure the particles would exert on a semipermeable boundary that passes solvent but not particles \citep{zb-12,chu2016active,johnson2021phase,wang2021vitrification}. Raising the number density increases both \(\Pi\) and the chemical potential \(\mu\) at fixed \(T\). Relative to perfectly hard spheres, finite particle hardness (``softness'') admits tiny overlaps/deformations that increase locally accessible free volume and the number of microstates; {for metastable states} at a given \(\phi\) this lowers \(\Pi\) and enhances metastability.

This trend appears in \textbf{Fig.~\ref{fig:fig_pi_6kt}}: our nearly-hard-sphere data, together with prior colloidal experiments \citep{prczcdo-96}, lie systematically \emph{below} hard-sphere liquid-state theory \citep{cs-69,hall1972another} and event-driven molecular dynamics (EDMD) for perfectly hard spheres \citep{pieprzyk2019thermodynamic} on the metastable branches. At coexistence, however, the same softness elevates the inferred coexistence pressure. Mechanistically, softness stabilizes amorphous (fluid-like) local structure within the coexistence region by adding short-range (vibrational) entropy, {reducing the needed gain in short-range entropy from ordered (crystalline) structure}; to satisfy \(\mu_{\mathrm{fluid}}=\mu_{\mathrm{solid}}\), a higher \(\Pi\) is then required than in the strict hard-sphere limit. Thus softness lowers metastable \(\Pi(\phi)\) but \emph{raises} \(\Pi_{\mathrm{coex}}\), reconciling the two observations.

Following Gispen \emph{et~al.}\ \citep{gispen2024finding}, the isothermal ``over-pressure''
\begin{equation*}
\Delta \Pi(\phi) \equiv \Pi(\phi)-\Pi_{\mathrm{coex}}
\end{equation*}
maps to the nucleation driving force via the Gibbs-Duhem relation \(d\mu = d\Pi/n\). They report a reduced \(\Delta \Pi\) on the solid branch, which they associate with proximity to a spinodal (superheat) limit \citep{wang2018homogeneous}, implying a smaller driving force for melting than for crystallization. This framework is consistent with the softness-induced shifts we observe in both metastable and coexistence pressures.

We speculate that the observed over-pressure condition explains why the coexistence line predicted for our softest particles (\(V_0=6kT\), \(B_2^*=0.985\)) in Fig.~\ref{fig:fig_pi_6kt} slopes downward at higher volume fractions toward the theoretical value. This behavior reflects an interplay between the over-pressure condition that increases with \(\phi\) and finite particle hardness. Stronger over-pressure favors sampling of periodic, ordered structures and greater short-range (vibrational) entropy, whereas finite hardness allows small deformations that increase configurational freedom and long-range entropy. For particles with \(V_0=6kT\) (\(B_2^*=0.985\)), at lower \(\phi\) (e.g., \(\phi=0.505\)) the over-pressure is weaker than the particles’ deformability, favoring retention of fluid-like structure and resulting in an overall higher coexistence pressure than hard-sphere theory, which predicts a larger crystal fraction. At higher volume fraction (e.g., \(\phi=0.54\)), the elevated over-pressure dominates finite-hardness effects, yielding a crystal fraction and pressure close to hard-sphere theory predictions. Consistent with Gispen \emph{et al.} \citep{gispen2024finding}, the system most closely approaches the theoretical coexistence tie line of Hoover and Ree \citep{hr-68} near the melting point.

This mechanistic perspective also clarifies the path dependence of the final state, illustrated in Fig.~\ref{fig:fig_xi_6kt}. Samples prepared near the metastable fluid line remain metastable for \(\phi \le 0.514\) due to weak over-pressure but phase separate for \(\phi \ge 0.515\). Conversely, samples prepared near the metastable crystal line experience strong over-pressure at low \(\phi\) and weak excess pressure at high \(\phi\); thus, the opposite trend arises: phase separation occurs readily for \(\phi \le 0.518\), whereas systems remain metastable crystals for \(\phi \ge 0.52\).

Considering finite-hardness particles diffusing within a fictitious enclosure, increasing hardness raises osmotic pressure; beyond a critical threshold, structural rearrangement into a crystalline phase reduces osmotic pressure and increases local entropy. To examine this relationship, we systematically increased hardness to \(V_0 = 15kT\), \(30kT\), and \(60kT\), corresponding to reduced second virial coefficients \(B_2^* = 0.990\), \(0.993\), and \(0.995\), respectively. Fig.~\ref{fig:fig_hardness}(a) shows that as particle hardness approaches the true hard-sphere limit, the predicted freezing and melting points converge toward theoretical values. For the freezing point, \(\phi_{\mathrm{F}} = 0.500\) (\(V_0 = 15kT\)), \(0.499\) (\(30kT\)), and \(0.497\) (\(60kT\)); for the melting point, \(\phi_{\mathrm{M}} = 0.544\), \(0.542\), and \(0.541\), respectively. Increased hardness also facilitated phase separation: with identical starting configurations, systems of harder particles exhibited phase coexistence across a broader portion of the theoretical coexistence region (see \S\ref{subsec:app_time}).

\begin{figure*}[!htbp]
	\vspace{0mm}
	\centering
	\includegraphics[width=0.75\linewidth]{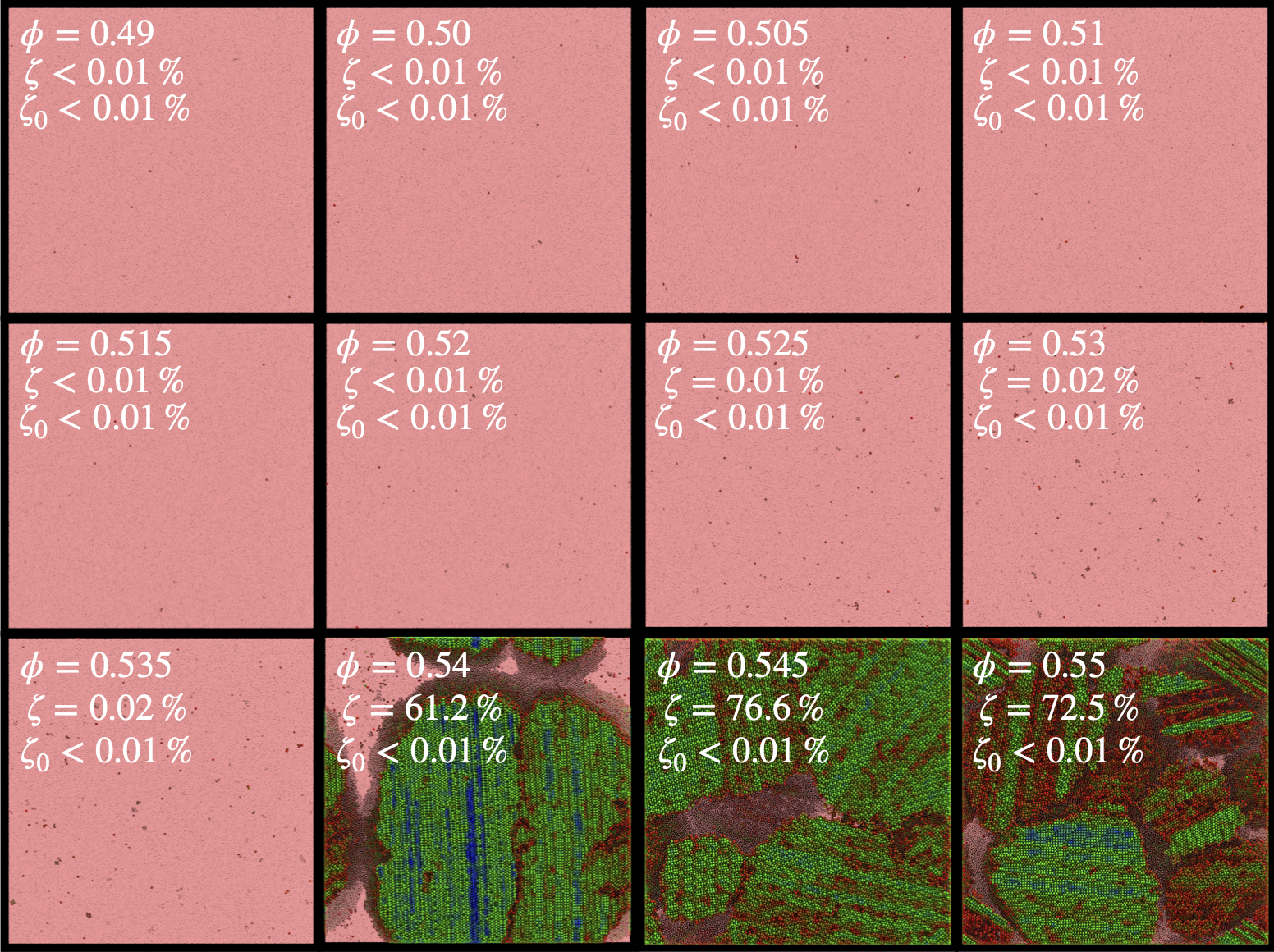}
	\caption{Phase behavior without crystal seeding. Simulation snapshots for samples with particles of hardness \(V_0 = 6kT\) (\(B_2^* = 0.985\)) initially on the metastable fluid line (\(\zeta_0 < 0.01\%\)), showing spontaneous phase separation after \(2{,}000\,a^2/D\). Particles are colored by the sixth-order average local-order parameter \(\bar{q}_6\), as in Fig.~\ref{fig:fig_image_panel}.}
	\label{fig:panel_0seed_6kT}
	\vspace{4mm}
	\centering
	\includegraphics[width=0.75\linewidth]{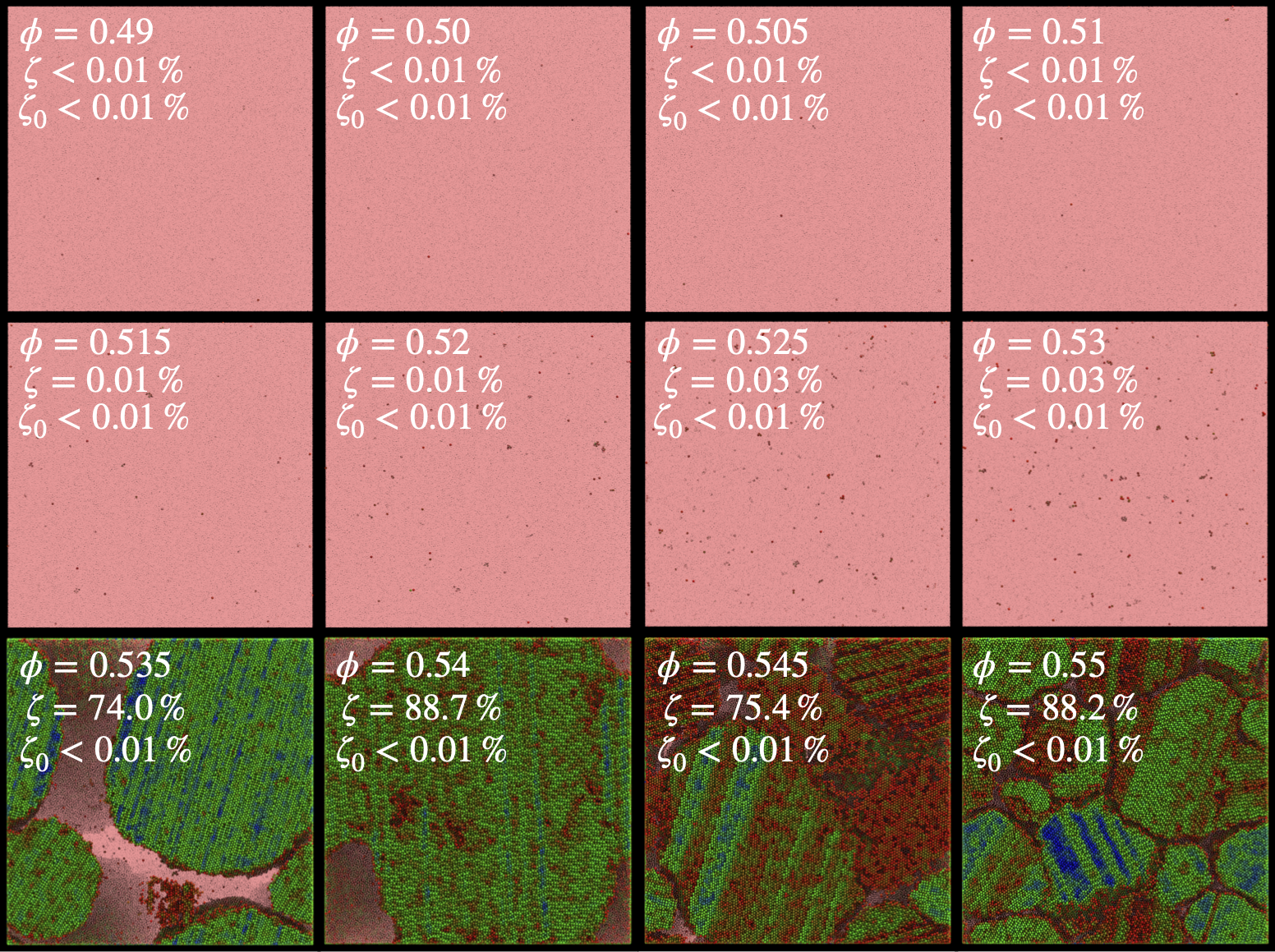}
	\caption{Phase behavior without crystal seeding. Simulation snapshots for samples with particles of hardness \(V_0 = 60kT\) (\(B_2^* = 0.995\)) initially on the metastable fluid line (\(\zeta_0 < 0.01\%\)), showing spontaneous phase separation after \(2{,}000\,a^2/D\). Particles are colored by the sixth-order average local-order parameter \(\bar{q}_6\), as in Fig.~\ref{fig:fig_image_panel}.}
	\label{fig:panel_0seed_60kT}
	\vspace{0mm}
\end{figure*}

Fig.~\ref{fig:fig_hardness}(b) presents the osmotic pressure for these systems. At each \(\phi\), samples were prepared near the fluid metastable line (as in Fig.~\ref{fig:fig_xi_6kt}), with corresponding initial crystal fractions shown in Fig.~\ref{fig:fig_image_panel}. Particle hardness was then set to one of the four values indicated, and simulation commenced. As expected, increased hardness raised the osmotic pressure of metastable states at all volume fractions, with the hardest samples approaching the purely hard-sphere values from hard-sphere liquid-state theory. As discussed above, greater hardness reduces amorphous configurations and long-range entropy, thereby promoting phase separation. This effect is particularly pronounced at low \(\phi\) (e.g., \(\phi = 0.505\)), where coexistence pressures approach theoretical predictions as hardness increases. The resulting coexistence line flattens and converges toward the Hoover-Ree theoretical tie line as the particles approach the truly hard-sphere limit.

In summary, simulations initiated with particle hardness \(V_0 = 6kT\) (\(B_2^* = 0.985\)) intended to represent very hard spheres and 2--4\% distributed crystal seeding produced phase separation consistent with prior studies but underpredicted metastable fluid and solid pressures. We attribute this discrepancy to finite particle softness. Increasing hardness toward the truly hard-sphere limit restored closer agreement with theoretical metastable and coexistence pressures and improved prediction of the phase envelope, yielding explicit phase separation in finite time. Hardening of already very hard particles increased the probability of phase separation, opposite the trend for soft particles modeled via WCA potential (\(B_2^* \lessapprox 0.8\)), where further softening lowers the phase envelope \citep{hermes2011nucleation,filion2011simulation,tateno2019influence,fiorucci2020effect,gispen2024finding}. For soft particles, rearrangements on the order of a particle diameter promote crystallization; in contrast, in the extremely hard-sphere regime, crystallization requires high pressures to drive short-range particle motion.

We compare these findings to the truly hard-sphere simulations of Pieprzyk \emph{et al.}~\citep{pieprzyk2019thermodynamic}, which did not show explicit phase separation but reproduced the hard-sphere liquid-state phase envelope and deduced the coexistence line from the chemical potential. Our extremely hard-sphere systems exhibited explicit phase separation in finite time but displayed slight over- and under-predictions driven by residual finite hardness. Two factors likely explain the differing outcomes. First, simulation methodology: for a given potential and ensemble, both EDMD and BD converge to the same Boltzmann distribution at equilibrium, and we do not suggest otherwise. Our point is kinetic: starting from a metastable fluid (or metastable crystal), the rate and pathway of phase separation (nucleation, growth, and coarsening) can differ significantly between EDMD and BD because the underlying dynamics are fundamentally different (ballistic, momentum-conserving versus overdamped, stochastic). As a result, whether phase separation is actually observed within accessible simulation times can differ between EDMD and BD, even at identical thermodynamic state points. Second, crystal seeding: distributed seeding in our BD simulations perturbs metastability and accelerates phase separation, whereas the absence of such perturbations in the EDMD study of Pieprzyk \emph{et al.} maintains metastability.

To isolate the effect of hardness, we next tested whether increasing \(V_0\) alone---without any crystal seeding---could destabilize the metastable state. This test provides a direct demonstration of Frenkel’s mechanistic model of phase separation in monodisperse, purely repulsive hard spheres, wherein short-range vibrational entropy increases with minimal perturbation of particle hardness. We focused on the very-hard regime to exclude mechanisms associated with soft-particle phase shifts \citep{hermes2011nucleation,filion2011simulation,tateno2019influence,fiorucci2020effect,gispen2024finding}.

Additional simulations with 0\% crystal seeding were therefore performed at the metastable lines, comparing phase behavior for very hard (\(6kT\), \(B_2^* = 0.985\)) and the hardest (\(60kT\), \(B_2^* = 0.995\)) spheres. Spontaneous phase separation occurred for \(\phi \ge 0.54\) in the \(6kT\) system (Fig.~\ref{fig:panel_0seed_6kT}) and for \(\phi \ge 0.535\) in the \(60kT\) system (Fig.~\ref{fig:panel_0seed_60kT}). Eliminating crystal seeding narrowed the observable coexistence window, but at fixed seeding fraction of zero, increasing hardness broadened this range. It is gratifying to observe spontaneous, equilibrium phase separation in a colloidal simulation with no crystal seeding, a nice demonstration of Frenkel’s proposed mechanism of entropy-driven phase separation in monodisperse, purely repulsive hard spheres, where short-range vibrational entropy is enhanced by minimally perturbed particle hardness.

\subsection{Time evolution of osmotic pressure and crystal fraction}
\label{subsec:app_time}
\begin{figure}[h]
	\vspace{-0mm}
	\centering
	\includegraphics[width=0.95\linewidth]{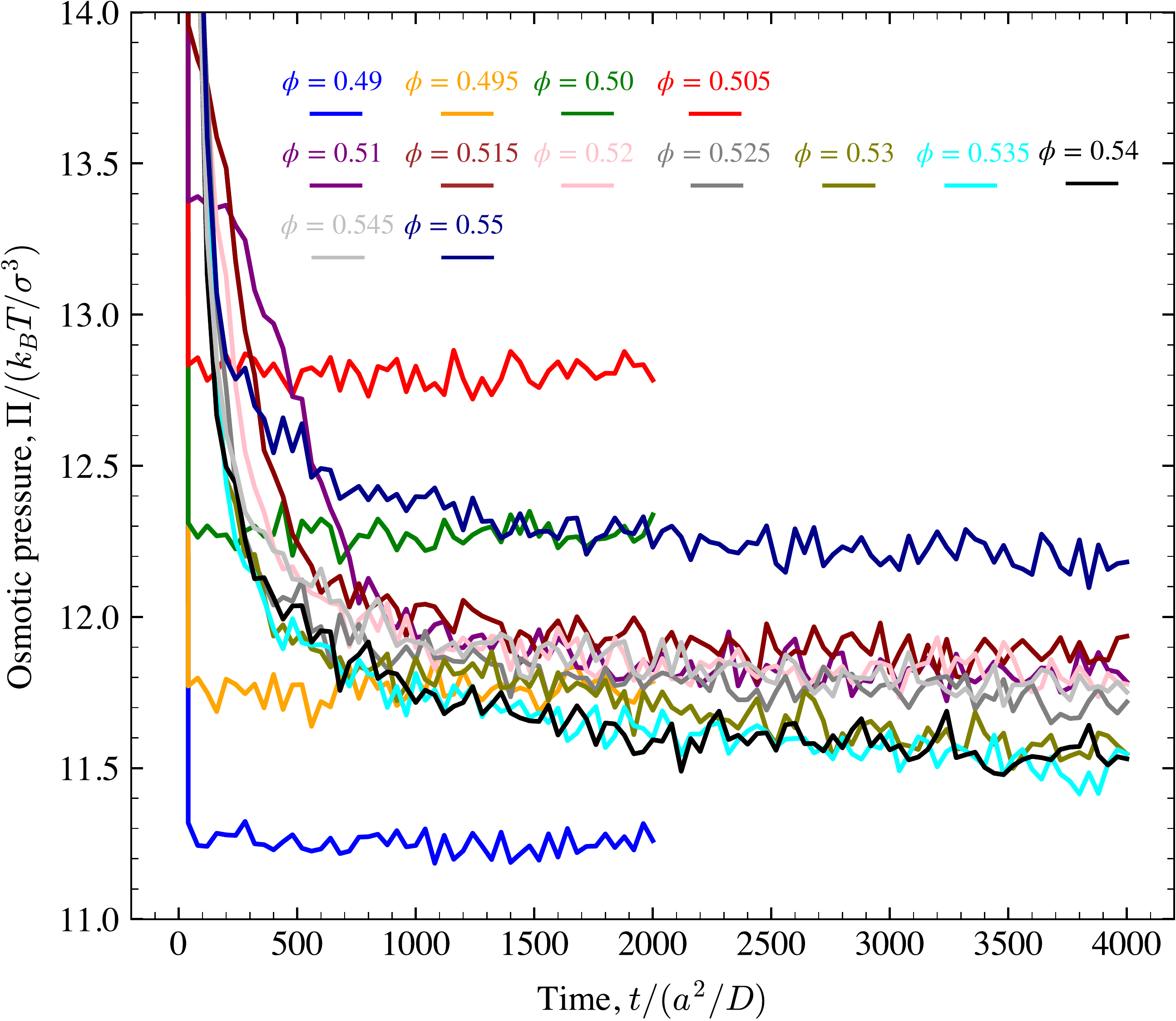}
	\caption{Time evolution of osmotic pressure for particles of hardness $V_0=60kT$ ($B_2^*=0.995$), over a duration of $4,000 a^2/D$. }
	\vspace{0mm}
	\label{fig:fig_OP_4000}
\end{figure}
\begin{figure*}[t]
	\vspace{0mm}
	\centering
	\includegraphics[width=0.9\textwidth]{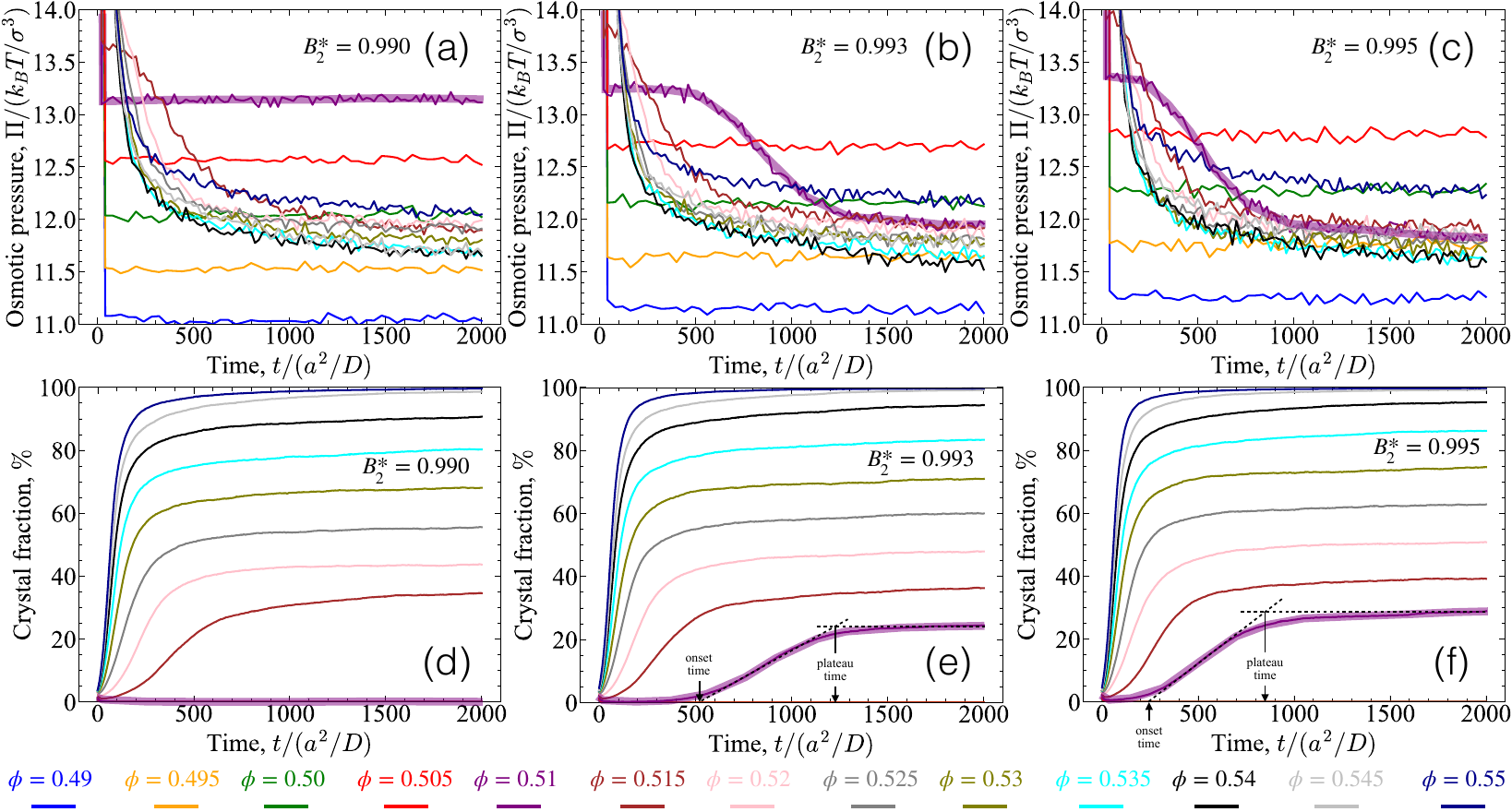}
	\caption{Temporal evolution of osmotic pressure and crystal fraction for $0.49 \le \phi \le 0.55$, as indicated in legend. Time is scaled on single-particle Brownian diffusion. Top row: Time evolution of osmotic pressure (scaled on the single-particle osmotic pressure). (a) $V_0=15kT$ ($B_2^*=0.990$), (b) $V_0=30kT$ ($B_2^*=0.993$) and (c) $V_0=60kT$ ($B_2^*=0.995$). Bottom row: crystal fraction, for (d) $V_0=15kT$ ($B_2^*=0.990$), (e) $V_0=30kT$ ($B_2^*=0.993$) and (f) $V_0=60kT$ ($B_2^*=0.995$).}
	\vspace{0mm}
	\label{fig:fig_OP_zeta}
\end{figure*}

Finite particle hardness evidently introduces kinetic effects during the transition from metastable states to the final phase separated state. We examined how these kinetics depend on particle hardness by monitoring the time evolution of osmotic pressure and crystal fraction for each hardness studied. Figure~\ref{fig:fig_OP_4000} shows the temporal evolution of pressure after initialization near the metastable line for our hardest particles (\(V_0 = 60kT\), \(B_2^* = 0.995\)). Each sample was monitored for at least \(2{,}000\,a^2/D\). Near the freezing point (\(0.49 \le \phi \le 0.505\)), the osmotic pressure decays almost instantaneously to its final equilibrium value, indicating that these initial configurations are already close to the final fluid state. For samples deeper into the coexistence and crystalline regions (\(0.51 \le \phi \le 0.55\)), where phase separation proceeds, the pressure exhibits an initial steep drop followed by a slower long-time decay. This two-stage relaxation is consistent with classical nucleation theory (CNT), which describes crystallization as a process overcoming an activation barrier \citep{CNT1926}. The activation process is most clearly manifested at \(\phi = 0.51\), where a brief shoulder in the curve indicates transient residence in the metastable state before rapid pressure decay as phase separation begins. Simulations extended to \(4{,}000\,a^2/D\) revealed no further change in the mean osmotic pressure, confirming that steady state was achieved by \(2{,}000\,a^2/D\).

We next investigated the influence of particle hardness on the temporal evolution of both osmotic pressure and crystal fraction, using samples prepared near the metastable line with minimal crystal seeding. Figure~\ref{fig:fig_OP_zeta} presents the evolution of osmotic pressure (top row) and crystal fraction (bottom row) for \(V_0 = 15kT\), \(30kT\), and \(60kT\). At each hardness, the long-time pressure plateau coincides with the saturation of the crystal fraction, consistent with CNT’s depiction of crystallization as an activated process.

Increasing particle hardness accelerates crystal growth. This trend is highlighted for \(\phi = 0.51\) in Fig.~\ref{fig:fig_OP_zeta} (bold curves). The onset of crystal nucleation is identified by the intersection of a tangent drawn through the initial growth regime with the time axis, while the completion time---corresponding to attainment of the equilibrium crystal fraction---is marked by the intersection between the growth-regime tangent and the long-time plateau tangent. Comparing panels (d), (e), and (f), both the onset and completion times decrease as particle hardness increases, confirming that systems of harder particles crystallize more readily. This observation aligns with the mechanistic picture outlined above: increasing hardness reduces available free volume, promoting crystalline order and enhancing short-range entropy, whereas reduced hardness expands free volume, providing additional configurations and greater long-range entropy. Harder particles also intensify the over-pressure driving force (cf.\ \S\ref{subsec:op}), thereby facilitating faster phase separation.

\section{Conclusions}
\label{sec:conclusions}

Entropically driven fluid--solid transitions in monodisperse, purely repulsive hard spheres (MPRHS) are firmly established: distinct phases, phase envelopes, and freezing and melting points have been predicted and observed across liquid-state theory, simulations, and experiments. The same framework implies a fluid--solid coexistence region in MPRHS that must be purely entropic in origin. Frenkel proposed a mechanistic basis for this in systems lacking polydispersity or anisotropy, arguing that crystallization trades long-range configurational entropy for short-range vibrational entropy~\citep{frenkel-93}. The present work shows how this entropy-exchange picture can be made operational in large-scale simulations that explicitly realize and interrogate phase behavior in nearly hard-sphere colloids.

Decades of simulations that nominally match the MPRHS model nonetheless struggle to produce spontaneous, long-lived fluid--crystal coexistence from unbiased homogeneous states: transient mixtures form but are typically overtaken by a single phase, and most studies quite appropriately focus on nucleation-rate measurements in a metastable fluid or crystal~\citep{tenwolde1996numerical,filion2010crystal,filion2011simulation,fiorucci2020effect,wohler2022hard,wangInreviewElusive}. Those nucleation studies have been essential for quantifying rates, interfacial free energies, and finite-size effects, and we view them as a cornerstone of the hard-sphere literature rather than something to be challenged~\citep{auer2001prediction,espinosa2013fluid,espinosa2014mold,espinosa2016seeding,montero2020young,montero2020interfacial,gispen2024finding}. Here we ask a complementary question. We emphasize that we do not question the existence or location of the hard-sphere coexistence window, nor do we claim to simulate a mathematically ideal MPRHS system. Instead, we treat the hard-sphere phase diagram as settled and pose a kinetic problem: how can one gently break metastability so that the coexistence state actually appears on accessible Brownian time scales, and how does Frenkel's configurational--vibrational entropy exchange operate once such minimal perturbations are present?

To address this, we computationally studied $2\times10^6$ nearly hard-sphere colloids undergoing Brownian dynamics with a short-ranged Morse interaction whose reduced second virial coefficient differs from the hard-sphere value by at most $ 1\%$. We then introduced two controlled perturbations: (i) tiny, distributed crystalline seeds at 2--4\% initial volume fraction, and (ii) small, explicitly quantified changes in particle hardness, scanning $B_2^* = 0.985, 0.990, 0.993,$ and $0.995$ (in contrast to much softer WCA-like models with $B_2^* \approx 0.73$~\citep{filion2011simulation,tateno2019influence,fiorucci2020effect,gispen2024finding}). With weak seeding at $B_2^* = 0.985$, we obtained explicit fluid--crystal phase separation and freezing/melting points $\phi_F \approx 0.497$ and $\phi_M \approx 0.541$, close to hard-sphere coexistence. As hardness was increased, the long-time crystal fraction and osmotic pressure moved systematically toward the hard-sphere fluid, crystal, and coexistence lines, indicating that our nearly hard-sphere system asymptotically recovers standard hard-sphere liquid-state behavior while retaining Brownian kinetics.

We then isolated the role of vibrational entropy by removing seeding and perturbing hardness alone. Samples prepared on the metastable fluid line with $B_2^* = 0.995$---only about $0.5\%$ below the hard-sphere limit---exhibited equilibrium fluid--crystal phase separation in finite time, whereas softer cases remained trapped in the metastable fluid. In our simulations, this minimal, quantified softness increased locally accessible free volume just enough to destabilize the metastable branch and allow coexistence to emerge. In this sense, our simulations provide a direct numerical realization of Frenkel's entropy-exchange mechanism in an MPRHS-like system: a small gain in short-range vibrational entropy enables the long-range configurational entropy loss required for crystallization, without materially altering the underlying hard-sphere thermodynamics.

Finally, our results have practical implications. They clarify how slight deviations from perfect hardness and small distributed seeds jointly control the dynamic accessibility of fluid--crystal coexistence in colloidal suspensions, and they illustrate how nearly hard-sphere Brownian simulations can be used to test and refine empirical freezing criteria (e.g., Hansen--Verlet and L\"{o}wen criteria) when crystallization is slow or difficult to observe directly. More broadly, they provide a concrete framework for connecting the well-established hard-sphere phase diagram to the specific microscopic perturbations and kinetic pathways that make the underlying entropy-exchange mechanism observable in realistic colloidal systems.

\begin{bmhead}[Acknowledgments]
The authors gratefully acknowledge very insightful feedback from an anonymous reviewer. JGW acknowledges helpful conversations with Dr. Gesse Roure.  The authors acknowledge the support of the National Science Foundation's computation resources: Anvil at the Purdue University's Rosen Center for Advanced Computing (RCAC) \citep{Anvil} and Ranch Storage at Texas Advanced Computing Center (TACC) at U.T. Austin through allocation CHM240060 from the ACCESS program \citep{ACCESS}, which is supported by U.S. National Science Foundation grants \#2138259, \#2138286, \#2138307, \#2137603, and \#2138296. Some of the computation for this work was also performed on the high performance computing infrastructure operated by Research Support Solutions in the Division of IT at the University of Missouri, Columbia MO DOI:https://doi.org/10.32469/10355/97710. We acknowledge computational support of the staff and resources of the University of Missouri's Hellbender High Performance Computing cluster.
\end{bmhead}

\begin{bmhead}[Declaration of Interests]
The authors report no conflict of interest.
\end{bmhead}

\begin{bmhead}[Funding Information]
The authors acknowledge the support of the National Science Foundation's computation resources: Anvil at the Purdue University's Rosen Center for Advanced Computing (RCAC) \citep{Anvil} and Ranch Storage at Texas Advanced Computing Center (TACC) at U.T. Austin through allocation CHM240060 from the ACCESS program \citep{ACCESS}, which is supported by U.S. National Science Foundation grants \#2138259, \#2138286, \#2138307, \#2137603, and \#2138296.
\end{bmhead}

\begin{bmhead}[Data Availability]
Data are stored on the Ranch Storage at Texas Advanced Computing Center (TACC) at U.T. Austin and are available upon request.
\end{bmhead}

\begin{bmhead}[Author Contributions]
\noindent J. Galen Wang --- Conceptualization (co-Lead); Data generation (Lead); Data analysis and curation (co-Lead); Methodology (Lead); Writing, original draft (co-Lead); Writing: review and editing (co-Lead); Funding acquisition (contributor).\\
\noindent Umesh Dhumal --- Conceptualization (contributor); Data generation (co-Lead); Data analysis and curation (co-Lead); Methodology (contributor); Writing, original draft (contributor); Writing: review and editing (contributor); Funding acquisition (contributor).\\
\noindent Monica E. A. Zakhari --- Conceptualization (contributor); Data generation (contributor); Data analysis and curation (contributor); Methodology (co-Lead); Writing, original draft (co-Lead); Writing: review and editing (contributor); Funding acquisition (contributor).\\
\noindent Roseanna N. Zia --- Conceptualization (co-Lead); Data generation (contributor); Data analysis and curation (co-Lead); Methodology (contributor); Investigation (Lead); Writing, original draft (co-Lead); Writing: review and editing (co-Lead); Funding acquisition (Lead).
\end{bmhead}

\balance

\clearpage
\begin{appen}
\section{A condensed history of the study of entropic phase transitions}
\textbf{1941: Kirkwood and Monroe}. Theory explains and predicts melting transition in purely repulsive hard sphere systems. Validated with experimental data with argon \citep{kirkwood1941statistical}. \\
\textbf{1949: Onsager}. First described purely entropic phase transitions arising from configurational entropy due to shape anisotropy \citep{onsager1949effects}.\\
\textbf{1957, 1959, 1960: Alder and Wainwright}. First EDMD simulations of perfectly hard, purely repulsive monodisperse hard spheres, showing phase transition but not coexistence. Authors called for larger simulations to explicitly show coexistence \citep{alder1957phase,alder1959studies, alder1960studies}.\\
\textbf{1957: International discussion led by Uhlenbeck}, in a letter edited by Percus in 1963 \citep{}. This discussion and letter addressed the seeming paradox of freezing in MPRHS systems that would seem to lead to a higher-entropy crystal. Demonstrated the mechanism was understood as real but mechanistically unexplained \citep{uhlenbeck1963p}.\\
\textbf{1968: Hoover and Ree}. Combination of theory and experiments produced the MPRHS phase envelope.  Lattice was used only for solid line. Virial equation (liquid-state theory) used to separately obtain the liquid line. Thermodynamics theory then  used to connect them to deduce the coexistence line\citep{hr-68}.\\
\textbf{1986: Pusey and van Megen}. Experimentally observed phase behavior in putatively hard-sphere colloids. Seminarl connection between colloidal and atomic phase behavior \citep{pvM-86}.\\
\textbf{1980s}: Extensive work demonstrating purely entropic phase transitions due to shape anisotropy and size polydispersity in colloids \citep{eppenga1984monte, camp1997phase, cuetos2007kinetic, cinacchi2010phase, miller2010crystallization, kallus2011dense, agarwal2011mesophase, haji2011phase, jiao2011communication, avendano2012phase, marechal2012freezing, peroukidis2013phase, dijkstra2014entropy, boles2016self, karas2019phase, lim2023engineering, kranendonk1991computer, bartlett1992superlattice, eldridge1993entropy, han1994freezing, dijkstra1998phase, dijkstra1999direct, bw-99, fs-03, zubarev2005condensation, zaccarelli2009crystallization, wilding2010phase, hopkins2011phase, filion2011self, dijkstra2014entropy, boles2016self, koshoji2021diverse, koshoji2021densest}.\\
\textbf{1993: Frenkel}. Proposed a mechanistic concept for phase transitions in MPRHS systems, the entropy exchange mechanism.  \citep{frenkel-93}. \\
\textbf{1996: ten Wolde and Frenkel \textit{et al.}}. Predicted that observing phase separation in large simulations of MPRHS would take 317,000,000 years due to the many configurations need to be sampled for the microstates to converge to the phase-separated macrostate \citep{tenwolde1996numerical}.\\
\textbf{2000: Frenkel}. Recast Onsager's phase transitions as a competition between translation versus orientational entropy. \citep{frenkel2000perspective}.\\
\textbf{2000s}: Extensive simulations studies of nucleation, nucleation rates, interfacial properties between liquid and crystal phases (many and inclusive citations in our article \citep{auer2001prediction, auer2004numerical, filion2010crystal, filion2011simulation, isobe2015hard, hermes2011nucleation, espinosa2016seeding, fiorucci2020effect, montero2020young, montero2020interfacial, gispen2024finding, ladd1977triple, davidchack1998simulation, noya2008determination, zykova2010monte, espinosa2013fluid, tateno2019influence, sanchez2021fcc, wohler2022hard}). Established understanding of why nucleation is much slower in simulations than experiments. Demonstration that seeding, gravity, direct construction of a crystal slab (`direct coexistence') and other triggers are needed to induce phase separation, otherwise either the liquid or the solid takes over the entire system. Use of algorithmic drivers in Monte Carlo simulations to establish phase diagrams.\\
\textbf{2019: Pieprzyk \textit{et al.}}. Simulated a pristine MPRHS system with 1,000,000 particles using EDMD (no suspending fluid) to recover phase transition. Produced a sharpened phase envelope. Produced a pure metastable liquid and a pure metastable crysta but not explicit phase separation. Reinforced the metastability of MPRHS. \citep{pieprzyk2019thermodynamic}.\\

\begin{figure}[h]
	\vspace{0mm}
	\centering
	\includegraphics[width=0.7\textwidth]{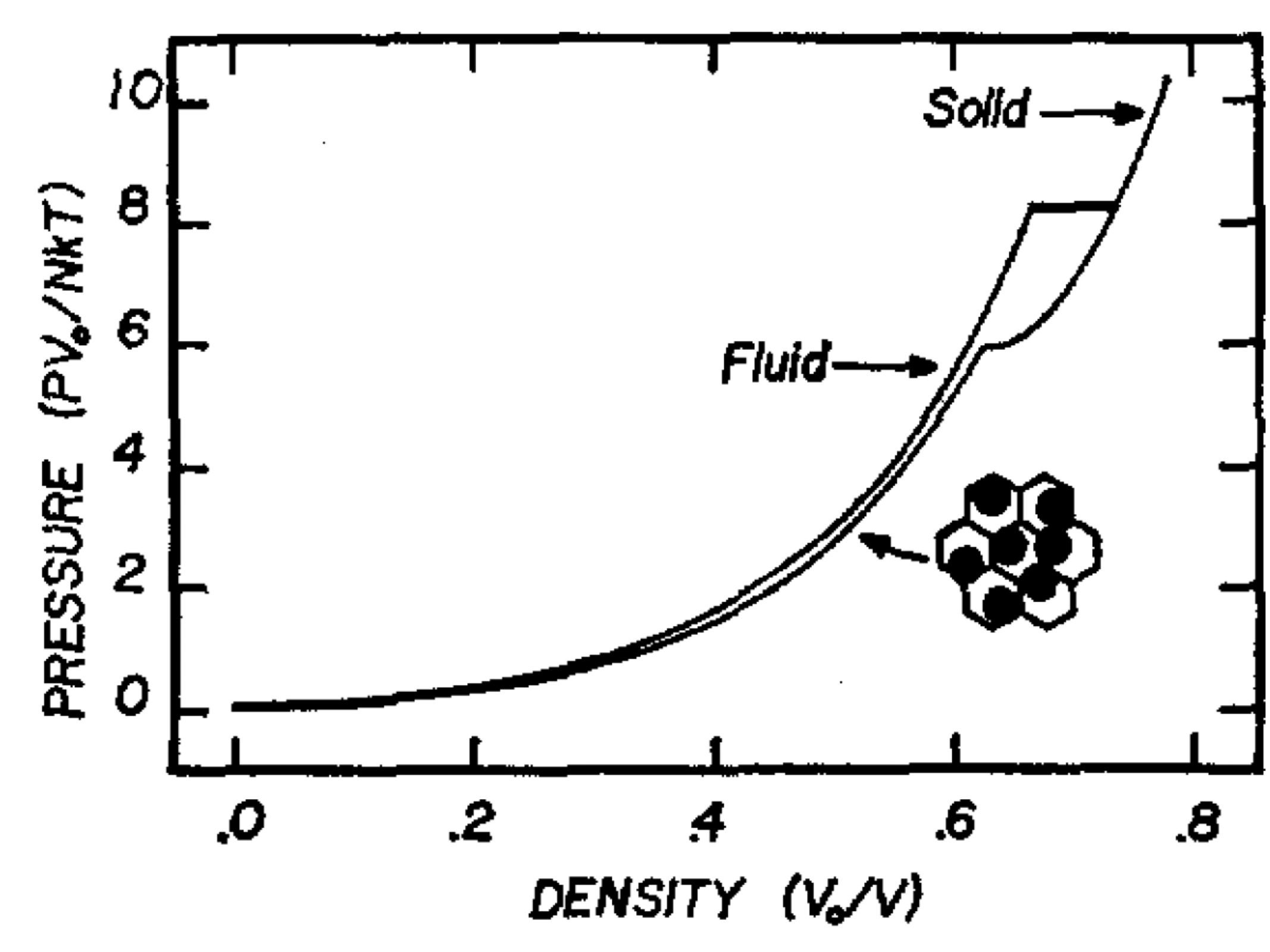}
	\caption{Hard-sphere equations of state. The fluid line is calculated theoretically from virial expansion, the solid line is obtained from single-occupancy lattice Monte Carlo simulation, and the tie line is calculated with equal-pressure and equal-chemical-potential conditions. Figure from \citep{hr-68}, with permission. }
	\vspace{0mm}
	\label{fig:fig_hr68}
\end{figure}

\section{Particle hardness in simulations}
\label{sec:app_hardness}
To model hard-sphere interactions, we use purely repulsive Morse potentials with a small range parameter $\kappa a=30$ (see Eq.~\ref{eqn:morse}) and plot the potential energy in Figure \ref{fig:fig_potential}. The Weeks-Chandler-Anderson (WCA) potential is another potential that is widely used in previous studies, with the following form:
\begin{equation}
	V(r) = 
\left\{
    \begin {aligned}
         & 4 V_0 \left[\left(\frac{\sigma}{r}\right)^{12}-\left(\frac{\sigma}{r}\right)^{6}+\frac{1}{4}\right], & \frac{r}{\sigma} < 2^{1/6} \\
         & 0, &\textrm{otherwise}.               
    \end{aligned}
\right.
	\label{eqn:wca}
\end{equation}
The $V_0=6kT$ Morse potential is quite hard, comparing to a $V_0=40kT$ WCA potential. We also make our particles much harder by increasing $V_0$ of Morse potential in Eq.~\ref{eqn:morse} and obtain a good model to represent hard spheres.

\section{Distribution of detailed crystalline structures}
\label{sec:app_c}
\begin{figure*}[h]
	\vspace{0mm}
	\centering
	\includegraphics[width=0.7\textwidth]{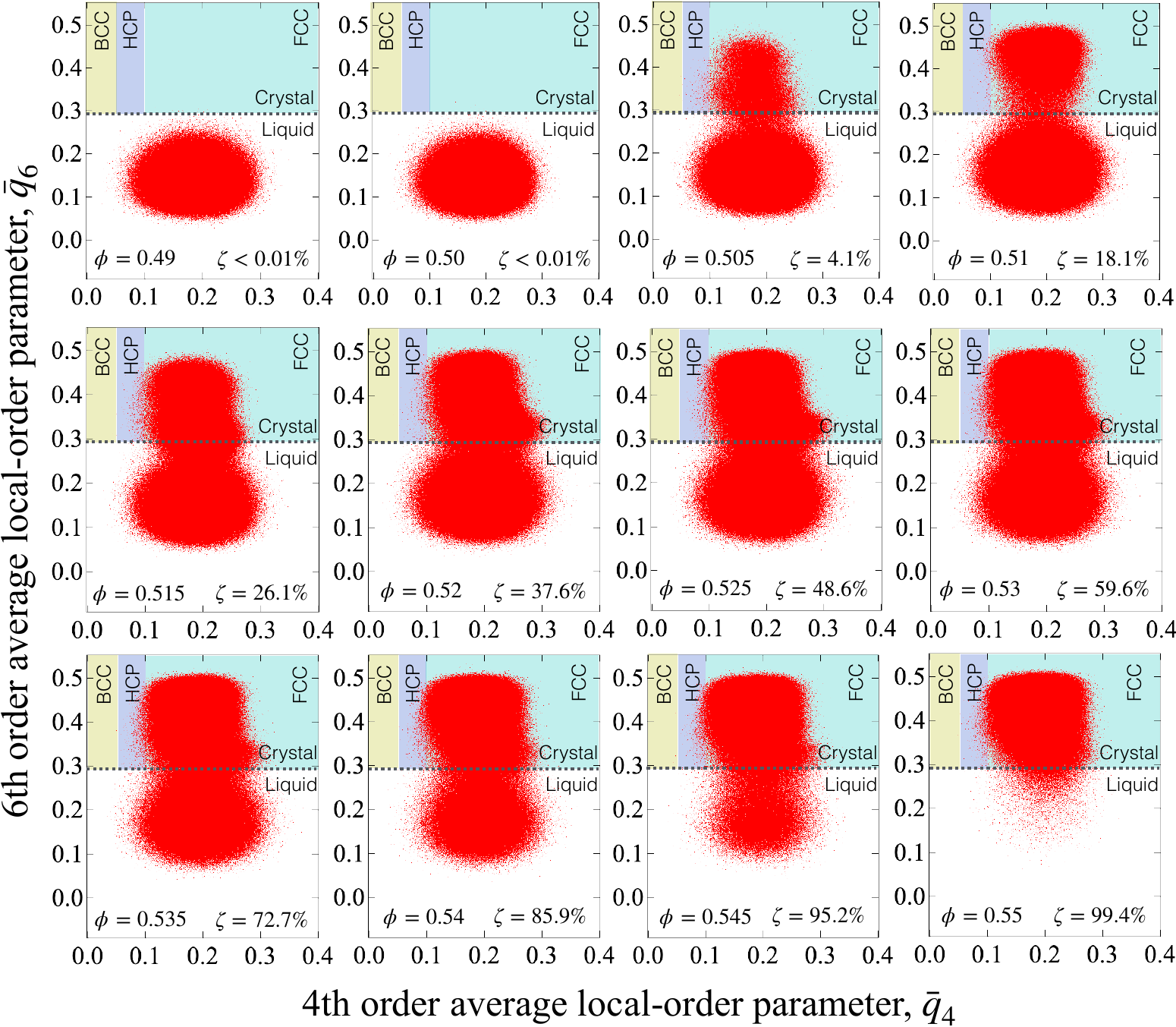}
	\caption{Per-particle $\bar{q}_6$ vs $\bar{q}_4$ plot for PRHS systems of $V_0=6kT$ particles at 12 volume fractions, as labeled in each plot. For all particles that are part of a crystalline structure ($\bar{q}_6 \ge 0.29$), the value of $\bar{q}_4$ determines the type of crystal structure \citep{Lechner2008, Kratzer2015}: BCC ($0 \le \bar{q}_4 \le 0.05$), HCP ($0.05<\bar{q}_4 \le 0.10$); and FCC ($\bar{q}_4>0.10$). A dotted line marks the boundary between fluid-like structure and crystalline structure. BCC, HCP, and FCC regions are marked and highlighted. }
	\vspace{0mm}
	\label{fig:fig_q6_vs_q4_6kt}
\end{figure*}

\begin{figure*}[h]
	\vspace{0mm}
	\centering
	\includegraphics[width=0.7\textwidth]{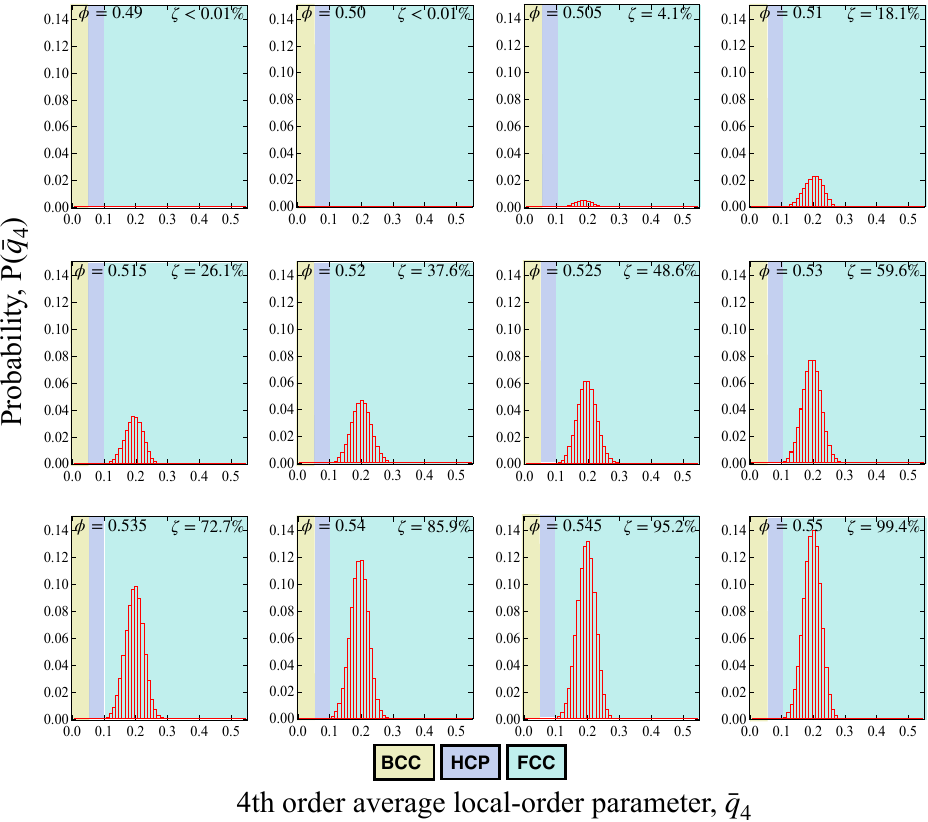}
	\caption{Probability P($\bar{q}_4$) vs $\bar{q}_4$ plot for systems of $V_0=6kT$ particles at 12 volume fractions at 2,000 BT, as labeled in each plot. For all particles that are crystalline ($\bar{q}_6 \ge 0.29$), BCC, HCP, and FCC regions are marked and highlighted the same way as Fig. \ref{fig:fig_q6_hist}.}
	\vspace{0mm}
	\label{fig:fig_q4_hist}
\end{figure*}

We quantify the detailed crystalline structure via the combined measurements of $\bar{q}_6$ and $\bar{q}_4$ [\textbf{Figure \ref{fig:fig_q6_vs_q4_6kt}}]. For crystalline structure ($\bar{q}_6 \ge 0.29$), values of $0< \bar{q}_4 \le 0.05$ signify BCC structure, $0.05 < \bar{q}_4 \le 0.1$ signifies HCP structure, and $\bar{q}_4>0.1$ signifies FCC structure. As labeled in the figure, $\bar{q}_4$ measurements reveal the structure of the coexisting crystalline state: nearly all crystalline regions are FCC. {We also quantify the statistics of crystalline structures via the histogram of $\bar{q}_4$ for all crystalline particles that have $\bar{q}_6 \ge 0.29$ [\textbf{Figure \ref{fig:fig_q4_hist}}]. There is only one peak of $P(\bar{q}_4)$ centered around $\bar{q}_4 \approx 0.2$, and it is clear that almost all crystalline colloids has $\bar{q}_4 > 0.1$. In fact, less than 0.1\% of structure is BCC or HCP. This distribution of crystalline structure is consistent with previous literature results, which indicate that FCC structure is slightly more stable than HCP structure \citep{frenkel1984new, Woodcock1997}.}

\end{appen}\clearpage


\begin{thebibliography}{159}
\expandafter\ifx\csname natexlab\endcsname\relax\def\natexlab#1{#1}\fi
\def\au#1{#1} \def\ed#1{#1} \def\yr#1{#1}\def\at#1{#1}\def\jt#1{\textit{#1}}
  \def\bt#1{#1}\def\bvol#1{\textbf{#1}} \def\vol#1{#1} \def\pg#1{#1}
  \def\publ#1{#1}\def\arxiv#1{#1}\def\org#1{#1}\def\st#1{\textit{#1}}

\bibitem[Abraham(2012)]{Abraham2012}
{\sc \au{Abraham, Farid}} \yr{2012} {\em {Homogeneous nucleation theory: the
  pretransition theory of vapor condensation}\/}, ,  \vol{vol.~1}.
  \publ{Elsevier}.

\bibitem[Ackerson(1993)]{ackerson1993order}
{\sc \au{Ackerson, Bruce~J}} \yr{1993}  \at{When order is disordered}.
  \jt{Nature}  \bvol{365}~(6441),  \pg{11--12}.

\bibitem[Agarwal \& Escobedo(2011)]{agarwal2011mesophase}
{\sc \au{Agarwal, Umang} \& \au{Escobedo, Fernando~A}} \yr{2011}  \at{Mesophase
  behaviour of polyhedral particles}.  \jt{Nature materials}  \bvol{10}~(3),
  \pg{230--235}.

\bibitem[Alder \& Wainwright(1959)]{alder1959studies}
{\sc \au{Alder, Berni~J} \& \au{Wainwright, Thomas~Everett}} \yr{1959}
  \at{Studies in molecular dynamics. i. general method}.  \jt{The Journal of
  Chemical Physics}  \bvol{31}~(2),  \pg{459--466}.

\bibitem[Alder \& Wainwright(1960)]{alder1960studies}
{\sc \au{Alder, Berni~Julian} \& \au{Wainwright, Thomas~Everett}} \yr{1960}
  \at{Studies in molecular dynamics. ii. behavior of a small number of elastic
  spheres}.  \jt{The Journal of Chemical Physics}  \bvol{33}~(5),
  \pg{1439--1451}.

\bibitem[Alder {\em et~al.\/}(1957)Alder, Wainwright {\em
  et~al.\/}]{alder1957phase}
{\sc \au{Alder, Berni~Julian}, \au{Wainwright, Thomas~Everett} \& \au{others}}
  \yr{1957}  \at{Phase transition for a hard sphere system}.  \jt{The Journal
  of Chemical Physics}  \bvol{27}~(5),  \pg{1208}.

\bibitem[Allen \& Tildesley(1987)]{at-87}
{\sc \au{Allen, M.~P.} \& \au{Tildesley, D.~J.}} \yr{1987} {\em Computer
  Simulation of Liquids\/}.  \publ{Oxford: Clarendon Press}.

\bibitem[Aponte-Rivera {\em et~al.\/}(2018)Aponte-Rivera, Su \&
  Zia]{aponte2018equilibrium}
{\sc \au{Aponte-Rivera, Christian}, \au{Su, Yu} \& \au{Zia, Roseanna~N}}
  \yr{2018}  \at{{Equilibrium structure and diffusion in concentrated
  hydrodynamically interacting suspensions confined by a spherical cavity}}.
  \jt{Journal of Fluid Mechanics}  \bvol{836},  \pg{413--450}.

\bibitem[Aponte-Rivera \& Zia(2016)]{aponte2016simulation}
{\sc \au{Aponte-Rivera, Christian} \& \au{Zia, Roseanna~N}} \yr{2016}
  \at{{Simulation of hydrodynamically interacting particles confined by a
  spherical cavity}}.  \jt{Physical Review Fluids}  \bvol{1}~(2),  \pg{023301}.

\bibitem[Aponte-Rivera \& Zia(2022)]{aponte2022confined}
{\sc \au{Aponte-Rivera, Christian} \& \au{Zia, Roseanna~N.}} \yr{2022}
  \at{{The confined Generalized Stokes-Einstein relation and its consequence on
  intracellular two-point microrheology}}.  \jt{Journal of Colloid and
  Interface Science}  \bvol{609},  \pg{423--433}.

\bibitem[Archer(2005)]{archer2005density}
{\sc \au{Archer, A.~J.}} \yr{2005}  \at{{Density functional theory for the
  freezing of soft-core fluids}}.  \jt{Physical Review E - Statistical,
  Nonlinear, and Soft Matter Physics}  \bvol{72}~(5),  \pg{1--7}.

\bibitem[Auer \& Frenkel(2001{\natexlab{{\em a\/}}})]{auer2001prediction}
{\sc \au{Auer, Stefan} \& \au{Frenkel, Daan}} \yr{2001{\natexlab{{\em a\/}}}}
  \at{Prediction of absolute crystal-nucleation rate in hard-sphere colloids}.
  \jt{Nature}  \bvol{409}~(6823),  \pg{1020--1023}.

\bibitem[Auer \& Frenkel(2001{\natexlab{{\em b\/}}})]{auer2001suppression}
{\sc \au{Auer, Stefan} \& \au{Frenkel, Daan}} \yr{2001{\natexlab{{\em b\/}}}}
  \at{Suppression of crystal nucleation in polydisperse colloids due to
  increase of the surface free energy}.  \jt{Nature}  \bvol{413}~(6857),
  \pg{711--713}.

\bibitem[Auer \& Frenkel(2004)]{auer2004numerical}
{\sc \au{Auer, Stefan} \& \au{Frenkel, Daan}} \yr{2004}  \at{Numerical
  prediction of absolute crystallization rates in hard-sphere colloids}.
  \jt{The Journal of Chemical Physics}  \bvol{120}~(6),  \pg{3015--3029}.

\bibitem[Avendano \& Escobedo(2012)]{avendano2012phase}
{\sc \au{Avendano, Carlos} \& \au{Escobedo, Fernando~A}} \yr{2012}  \at{Phase
  behavior of rounded hard-squares}.  \jt{Soft Matter}  \bvol{8}~(17),
  \pg{4675--4681}.

\bibitem[Balescu(1975)]{balescu1975equilibrium}
{\sc \au{Balescu, R.}} \yr{1975} {\em Equilibrium and Non-Equilibrium
  Statistical Mechanics\/}. {\em A Wiley interscience publication\/} .
  \publ{Wiley}.

\bibitem[Bannerman {\em et~al.\/}(2010)Bannerman, Lue \&
  Woodcock]{bannerman2010thermodynamic}
{\sc \au{Bannerman, Marcus~N}, \au{Lue, Leo} \& \au{Woodcock, Leslie~V}}
  \yr{2010}  \at{Thermodynamic pressures for hard spheres and closed-virial
  equation-of-state}.  \jt{The Journal of Chemical Physics}  \bvol{132}~(8),
  \pg{084507}.

\bibitem[Bartlett {\em et~al.\/}(1992)Bartlett, Ottewill \&
  Pusey]{bartlett1992superlattice}
{\sc \au{Bartlett, P}, \au{Ottewill, RH} \& \au{Pusey, PN}} \yr{1992}
  \at{Superlattice formation in binary mixtures of hard-sphere colloids}.
  \jt{Physical Review Letters}  \bvol{68}~(25),  \pg{3801}.

\bibitem[Bartlett \& Warren(1999)]{bw-99}
{\sc \au{Bartlett, Paul} \& \au{Warren, Patrick~B.}} \yr{1999}  \at{Reentrant
  melting in polydispersed hard spheres}.  \jt{Phys. Rev. Lett.}  \bvol{82},
  \pg{1979--1982}.

\bibitem[Batchelor(1977)]{batchelor1977effect}
{\sc \au{Batchelor, George~K}} \yr{1977}  \at{The effect of brownian motion on
  the bulk stress in a suspension of spherical particles}.  \jt{Journal of
  fluid mechanics}  \bvol{83}~(1),  \pg{97--117}.

\bibitem[Bergenholtz {\em et~al.\/}(2002)Bergenholtz, Brady \& Vicic]{bbv-02}
{\sc \au{Bergenholtz, J.}, \au{Brady, J.~F.} \& \au{Vicic, M.}} \yr{2002}
  \at{The non-newtonian rheology of dilute colloidal suspensions}.  \jt{J.
  Fluid Mech.}  \bvol{456},  \pg{239–275}.

\bibitem[Boerner {\em et~al.\/}(2023)Boerner, Deems, Furlani, Knuth \&
  Towns]{ACCESS}
{\sc \au{Boerner, Timothy~J.}, \au{Deems, Stephen}, \au{Furlani, Thomas~R.},
  \au{Knuth, Shelley~L.} \& \au{Towns, John}} \yr{2023} Access: Advancing
  innovation: Nsf's advanced cyberinfrastructure coordination ecosystem:
  Services \& support.  \bt{In {\em Practice and Experience in Advanced
  Research Computing 2023: Computing for the Common Good\/}}, {\em PEARC '23\/}
  ,  \pg{p. 173–176}.  \publ{New York, NY, USA: Association for Computing
  Machinery}.

\bibitem[Boles {\em et~al.\/}(2016)Boles, Engel \& Talapin]{boles2016self}
{\sc \au{Boles, Michael~A}, \au{Engel, Michael} \& \au{Talapin, Dmitri~V}}
  \yr{2016}  \at{Self-assembly of colloidal nanocrystals: from intricate
  structures to functional materials}.  \jt{Chemical reviews}  \bvol{116}~(18),
   \pg{11220--11289}.

\bibitem[Brady(1993)]{brady1993brownian}
{\sc \au{Brady, John~F}} \yr{1993}  \at{Brownian motion, hydrodynamics, and the
  osmotic pressure}.  \jt{The Journal of Chemical Physics}  \bvol{98}~(4),
  \pg{3335--3341}.

\bibitem[Br{\"u}nger {\em et~al.\/}(1984)Br{\"u}nger, Brooks \&
  Karplus]{bbk-84}
{\sc \au{Br{\"u}nger, Axel}, \au{Brooks, Charles~L.} \& \au{Karplus, Martin}}
  \yr{1984}  \at{Stochastic boundary conditions for molecular dynamics
  simulations of st2 water}.  \jt{Chem. Phys. Lett.}  \bvol{105}~(5),  \pg{495
  -- 500}.

\bibitem[Bryant {\em et~al.\/}(2002)Bryant, Williams, Qian, Snook, Perez \&
  Pincet]{bryant2002hard}
{\sc \au{Bryant, Gary}, \au{Williams, Stephen~R}, \au{Qian, L}, \au{Snook, IK},
  \au{Perez, E} \& \au{Pincet, F}} \yr{2002}  \at{How hard is a colloidal
  ``hard-sphere'' interaction?}  \jt{Physical Review E}  \bvol{66}~(6),
  \pg{060501}.

\bibitem[Camp \& Allen(1997)]{camp1997phase}
{\sc \au{Camp, Philip~J} \& \au{Allen, Michael~P}} \yr{1997}  \at{Phase diagram
  of the hard biaxial ellipsoid fluid}.  \jt{The Journal of Chemical Physics}
  \bvol{106}~(16),  \pg{6681--6688}.

\bibitem[Carnahan \& Starling(1969)]{cs-69}
{\sc \au{Carnahan, Norman~F.} \& \au{Starling, Kenneth~E.}} \yr{1969}
  \at{Equation of state for nonattracting rigid spheres}.  \jt{J. Chem. Phys.}
  \bvol{51}~(2),  \pg{635--636}.

\bibitem[Castelletto {\em et~al.\/}(2002)Castelletto, Caillet, Hamley \&
  Yang]{castelletto2002liquid}
{\sc \au{Castelletto, V.}, \au{Caillet, C.}, \au{Hamley, I.~W.} \& \au{Yang,
  Z.}} \yr{2002}  \at{{Liquid-solid transition in a model hard sphere system of
  block copolymer micelles}}.  \jt{Physical Review E - Statistical Physics,
  Plasmas, Fluids, and Related Interdisciplinary Topics}  \bvol{65}~(5),
  \pg{4}.

\bibitem[Chu \& Zia(2016)]{chu2016active}
{\sc \au{Chu, Henry~CW} \& \au{Zia, Roseanna~N}} \yr{2016}  \at{Active
  microrheology of hydrodynamically interacting colloids: Normal stresses and
  entropic energy density}.  \jt{Journal of Rheology}  \bvol{60}~(4),
  \pg{755--781}.

\bibitem[Cinacchi \& van Duijneveldt(2010)]{cinacchi2010phase}
{\sc \au{Cinacchi, Giorgio} \& \au{van Duijneveldt, Jeroen~S}} \yr{2010}
  \at{Phase behavior of contact lens-like particles: Entropy-driven competition
  between isotropic- nematic phase separation and clustering}.  \jt{The Journal
  of Physical Chemistry Letters}  \bvol{1}~(4),  \pg{787--791}.

\bibitem[Clisby \& McCoy(2006)]{clisby2006ninth}
{\sc \au{Clisby, Nathan} \& \au{McCoy, Barry~M}} \yr{2006}  \at{Ninth and tenth
  order virial coefficients for hard spheres in d dimensions}.  \jt{Journal of
  Statistical Physics}  \bvol{122}~(1),  \pg{15--57}.

\bibitem[Cuetos \& Dijkstra(2007)]{cuetos2007kinetic}
{\sc \au{Cuetos, Alejandro} \& \au{Dijkstra, Marjolein}} \yr{2007}  \at{Kinetic
  pathways for the isotropic-nematic phase transition in a system of colloidal
  hard rods: a simulation study}.  \jt{Physical Review Letters}  \bvol{98}~(9),
   \pg{095701}.

\bibitem[Davidchack \& Laird(1998)]{davidchack1998simulation}
{\sc \au{Davidchack, Ruslan~L} \& \au{Laird, Brian~B}} \yr{1998}
  \at{Simulation of the hard-sphere crystal--melt interface}.  \jt{The Journal
  of Chemical Physics}  \bvol{108}~(22),  \pg{9452--9462}.

\bibitem[Debenedetti(1996)]{debenedetti-96}
{\sc \au{Debenedetti, Pablo~G.}} \yr{1996} {\em Metastable liquids: Concepts
  and Principles\/}.  \publ{Princeton University Press}.

\bibitem[Di {\em et~al.\/}(2014)Di, Peng \& McKenna]{dpm-14}
{\sc \au{Di, X.}, \au{Peng, X.} \& \au{McKenna, G.~B.}} \yr{2014}
  \at{{Dynamics of a thermo-responsive microgel colloid near to the glass
  transition}}.  \jt{J. Chem. Phys.}  \bvol{140}~(5),  \pg{054903}.

\bibitem[Di {\em et~al.\/}(2011)Di, Win, McKenna, Narita, Lequeux, Pullela \&
  Cheng]{dwmnlpc-11}
{\sc \au{Di, Xiaojun}, \au{Win, K.~Z.}, \au{McKenna, Gregory~B.}, \au{Narita,
  Tetsuharu}, \au{Lequeux, Fran{\ifmmode \mbox{\c{c}}\else \c{c}\fi}ois},
  \au{Pullela, Srinivasa~Rao} \& \au{Cheng, Zhengdong}} \yr{2011}
  \at{Signatures of structural recovery in colloidal glasses}.  \jt{Phys. Rev.
  Lett.}  \bvol{106},  \pg{095701}.

\bibitem[Dijkstra(2014)]{dijkstra2014entropy}
{\sc \au{Dijkstra, Marjolein}} \yr{2014}  \at{Entropy-driven phase transitions
  in colloids: From spheres to anisotropic particles}.  \jt{Advances in
  Chemical Physics}  \bvol{156},  \pg{35--71}.

\bibitem[Dijkstra {\em et~al.\/}(1998)Dijkstra, van Roij \&
  Evans]{dijkstra1998phase}
{\sc \au{Dijkstra, Marjolein}, \au{van Roij, Ren{\'e}} \& \au{Evans, Robert}}
  \yr{1998}  \at{Phase behavior and structure of binary hard-sphere mixtures}.
  \jt{Physical Review Letters}  \bvol{81}~(11),  \pg{2268}.

\bibitem[Dijkstra {\em et~al.\/}(1999)Dijkstra, van Roij \&
  Evans]{dijkstra1999direct}
{\sc \au{Dijkstra, Marjolein}, \au{van Roij, Ren{\'e}} \& \au{Evans, Robert}}
  \yr{1999}  \at{Direct simulation of the phase behavior of binary hard-sphere
  mixtures: Test of the depletion potential description}.  \jt{Physical Review
  Letters}  \bvol{82}~(1),  \pg{117}.

\bibitem[Eldridge {\em et~al.\/}(1993)Eldridge, Madden \&
  Frenkel]{eldridge1993entropy}
{\sc \au{Eldridge, MD}, \au{Madden, PA} \& \au{Frenkel, D}} \yr{1993}
  \at{Entropy-driven formation of a superlattice in a hard-sphere binary
  mixture}.  \jt{Nature}  \bvol{365}~(6441),  \pg{35--37}.

\bibitem[Eppenga \& Frenkel(1984)]{eppenga1984monte}
{\sc \au{Eppenga, R} \& \au{Frenkel, Daan}} \yr{1984}  \at{Monte carlo study of
  the isotropic and nematic phases of infinitely thin hard platelets}.
  \jt{Molecular physics}  \bvol{52}~(6),  \pg{1303--1334}.

\bibitem[Erigi {\em et~al.\/}(2023)Erigi, Dhumal \& Tripathy]{erigi2023phase}
{\sc \au{Erigi, Umashankar}, \au{Dhumal, Umesh} \& \au{Tripathy, Mukta}}
  \yr{2023}  \at{Phase behavior of mixtures of hard colloids and soft
  coarse-grained macromolecules}.  \jt{The Journal of Chemical Physics}
  \bvol{159}~(16),  \pg{164901}.

\bibitem[Espinosa {\em et~al.\/}(2014)Espinosa, Vega \& Sanz]{espinosa2014mold}
{\sc \au{Espinosa, JR}, \au{Vega, C} \& \au{Sanz, E}} \yr{2014}  \at{The mold
  integration method for the calculation of the crystal-fluid interfacial free
  energy from simulations}.  \jt{The Journal of Chemical Physics}
  \bvol{141}~(13),  \pg{134709}.

\bibitem[Espinosa {\em et~al.\/}(2013)Espinosa, Sanz, Valeriani \&
  Vega]{espinosa2013fluid}
{\sc \au{Espinosa, Jorge~R}, \au{Sanz, Eduardo}, \au{Valeriani, Chantal} \&
  \au{Vega, Carlos}} \yr{2013}  \at{On fluid-solid direct coexistence
  simulations: The pseudo-hard sphere model}.  \jt{The Journal of Chemical
  Physics}  \bvol{139}~(14),  \pg{144502}.

\bibitem[Espinosa {\em et~al.\/}(2016)Espinosa, Vega, Valeriani \&
  Sanz]{espinosa2016seeding}
{\sc \au{Espinosa, Jorge~R}, \au{Vega, Carlos}, \au{Valeriani, Chantal} \&
  \au{Sanz, Eduardo}} \yr{2016}  \at{Seeding approach to crystal nucleation}.
  \jt{The Journal of Chemical Physics}  \bvol{144}~(3),  \pg{034501}.

\bibitem[Fasolo \& Sollich(2003)]{fs-03}
{\sc \au{Fasolo, Moreno} \& \au{Sollich, Peter}} \yr{2003}  \at{Equilibrium
  phase behavior of polydisperse hard spheres}.  \jt{Phys. Rev. Lett.}
  \bvol{91},  \pg{068301}.

\bibitem[Fern{\'a}ndez {\em et~al.\/}(2012)Fern{\'a}ndez, Martin-Mayor, Seoane
  \& Verrocchio]{fernandez2012equilibrium}
{\sc \au{Fern{\'a}ndez, LA}, \au{Martin-Mayor, V}, \au{Seoane, B} \&
  \au{Verrocchio, Paolo}} \yr{2012}  \at{Equilibrium fluid-solid coexistence of
  hard spheres}.  \jt{Physical Review Letters}  \bvol{108}~(16),  \pg{165701}.

\bibitem[Filion {\em et~al.\/}(2010)Filion, Hermes, Ni \&
  Dijkstra]{filion2010crystal}
{\sc \au{Filion, Laura}, \au{Hermes, Michiel}, \au{Ni, Ran} \& \au{Dijkstra,
  Marjolein}} \yr{2010}  \at{Crystal nucleation of hard spheres using molecular
  dynamics, umbrella sampling, and forward flux sampling: A comparison of
  simulation techniques}.  \jt{The Journal of Chemical Physics}
  \bvol{133}~(24),  \pg{244115}.

\bibitem[Filion {\em et~al.\/}(2011{\natexlab{{\em a\/}}})Filion, Hermes, Ni,
  Vermolen, Kuijk, Christova, Stiefelhagen, Vissers, Van~Blaaderen \&
  Dijkstra]{filion2011self}
{\sc \au{Filion, L}, \au{Hermes, M}, \au{Ni, R}, \au{Vermolen, ECM}, \au{Kuijk,
  A}, \au{Christova, CG}, \au{Stiefelhagen, JCP}, \au{Vissers, T},
  \au{Van~Blaaderen, A} \& \au{Dijkstra, M}} \yr{2011{\natexlab{{\em a\/}}}}
  \at{Self-assembly of a colloidal interstitial solid with tunable sublattice
  doping}.  \jt{Physical Review Letters}  \bvol{107}~(16),  \pg{168302}.

\bibitem[Filion {\em et~al.\/}(2011{\natexlab{{\em b\/}}})Filion, Ni, Frenkel
  \& Dijkstra]{filion2011simulation}
{\sc \au{Filion, Laura}, \au{Ni, Ran}, \au{Frenkel, Daan} \& \au{Dijkstra,
  Marjolein}} \yr{2011{\natexlab{{\em b\/}}}}  \at{Simulation of nucleation in
  almost hard-sphere colloids: The discrepancy between experiment and
  simulation persists}.  \jt{The Journal of Chemical Physics}  \bvol{134}~(13),
   \pg{134901}.

\bibitem[Fiorucci {\em et~al.\/}(2020)Fiorucci, Coli, Padding \&
  Dijkstra]{fiorucci2020effect}
{\sc \au{Fiorucci, Giulia}, \au{Coli, Gabriele~M}, \au{Padding, Johan~T} \&
  \au{Dijkstra, Marjolein}} \yr{2020}  \at{The effect of hydrodynamics on the
  crystal nucleation of nearly hard spheres}.  \jt{The Journal of Chemical
  Physics}  \bvol{152}~(6),  \pg{064903}.

\bibitem[Foss \& Brady(2000)]{fb-00}
{\sc \au{Foss, David~R.} \& \au{Brady, John~F.}} \yr{2000}  \at{Brownian
  dynamics simulation of hard-sphere colloidal dispersions}.  \jt{J. Rheol.}
  \bvol{44}~(3),  \pg{629--651}.

\bibitem[Frenkel(1993)]{frenkel-93}
{\sc \au{Frenkel, Daan}} \yr{1993}  \at{Order through disorder: entropy strikes
  back}.  \jt{Phys. World}  \bvol{6}~(2),  \pg{24--25}.

\bibitem[Frenkel(2000)]{frenkel2000perspective}
{\sc \au{Frenkel, Daan}} \yr{2000}  \at{{Perspective on “The effect of shape
  on the interaction of colloidal particles” Onsager L (1949) Ann NY Acad Sci
  51: 627}}.  \jt{Theoretical Chemistry Accounts}  \bvol{103}~(3),
  \pg{212--213}.

\bibitem[Frenkel \& Ladd(1984)]{frenkel1984new}
{\sc \au{Frenkel, Daan} \& \au{Ladd, Anthony~JC}} \yr{1984}  \at{New monte
  carlo method to compute the free energy of arbitrary solids. application to
  the fcc and hcp phases of hard spheres}.  \jt{The Journal of Chemical
  Physics}  \bvol{81}~(7),  \pg{3188--3193}.

\bibitem[Frenkel \& Smit(2002)]{frenkel2002understanding}
{\sc \au{Frenkel, Daan} \& \au{Smit, Berend}} \yr{2002} {\em Understanding
  molecular simulation: from algorithms to applications\/}.  \publ{Elsevier}.

\bibitem[Gispen \& Dijkstra(2024)]{gispen2024finding}
{\sc \au{Gispen, Willem} \& \au{Dijkstra, Marjolein}} \yr{2024}  \at{Finding
  the differences: Classical nucleation perspective on homogeneous melting and
  freezing of hard spheres}.  \jt{The Journal of Chemical Physics}
  \bvol{160}~(14),  \pg{141102}.

\bibitem[Gonzalez {\em et~al.\/}(2021)Gonzalez, Aponte-Rivera \&
  Zia]{gonzalez2021impact}
{\sc \au{Gonzalez, Emma}, \au{Aponte-Rivera, Christian} \& \au{Zia,
  Roseanna~N.}} \yr{2021}  \at{{Impact of polydispersity and confinement on
  diffusion in hydrodynamically interacting colloidal suspensions}}.
  \jt{Journal of Fluid Mechanics}  \bvol{925},  \pg{A35}.

\bibitem[Gupta {\em et~al.\/}(2015)Gupta, Camargo, Stellbrink, Allgaier,
  Radulescu, Lindner, Zaccarelli, Likos \& Richter]{gupta2015dynamic}
{\sc \au{Gupta, Sudipta}, \au{Camargo, Manuel}, \au{Stellbrink, J{\"o}rg},
  \au{Allgaier, J{\"u}rgen}, \au{Radulescu, Aurel}, \au{Lindner, Peter},
  \au{Zaccarelli, Emanuela}, \au{Likos, Christos~N} \& \au{Richter, Dieter}}
  \yr{2015}  \at{Dynamic phase diagram of soft nanocolloids}.  \jt{Nanoscale}
  \bvol{7}~(33),  \pg{13924--13934}.

\bibitem[Haji-Akbari {\em et~al.\/}(2011)Haji-Akbari, Engel \&
  Glotzer]{haji2011phase}
{\sc \au{Haji-Akbari, Amir}, \au{Engel, Michael} \& \au{Glotzer, Sharon~C}}
  \yr{2011}  \at{Phase diagram of hard tetrahedra}.  \jt{The Journal of
  Chemical Physics}  \bvol{135}~(19),  \pg{194101}.

\bibitem[Hall(1972)]{hall1972another}
{\sc \au{Hall, Kenneth~R}} \yr{1972}  \at{Another hard-sphere equation of
  state}.  \jt{The Journal of Chemical Physics}  \bvol{57}~(6),
  \pg{2252--2254}.

\bibitem[Han \& Herzfeld(1994)]{han1994freezing}
{\sc \au{Han, J} \& \au{Herzfeld, J}} \yr{1994}  \at{The freezing transition of
  bidisperse hard spheres: a simple model}.  \jt{Molecular physics}
  \bvol{82}~(3),  \pg{617--628}.

\bibitem[Hermes {\em et~al.\/}(2011)Hermes, Vermolen, Leunissen, Vossen,
  Van~Oostrum, Dijkstra \& Van~Blaaderen]{hermes2011nucleation}
{\sc \au{Hermes, M}, \au{Vermolen, ECM}, \au{Leunissen, ME}, \au{Vossen, DLJ},
  \au{Van~Oostrum, PDJ}, \au{Dijkstra, M} \& \au{Van~Blaaderen, A}} \yr{2011}
  \at{Nucleation of colloidal crystals on configurable seed structures}.
  \jt{Soft Matter}  \bvol{7}~(10),  \pg{4623--4628}.

\bibitem[Hernández-Guzmán \& Weeks(2009)]{hw-09}
{\sc \au{Hernández-Guzmán, Jessica} \& \au{Weeks, Eric~R.}} \yr{2009}
  \at{The equilibrium intrinsic crystal–liquid interface of colloids}.
  \jt{Proc. Nat. Acad. Sci.}  \bvol{106}~(36),  \pg{15198--15202},
  \arxiv{arXiv: http://www.pnas.org/content/106/36/15198.full.pdf}.

\bibitem[Heyes \& Melrose(1993)]{hm-93}
{\sc \au{Heyes, D.~M.} \& \au{Melrose, J.~R.}} \yr{1993}  \at{{B}rownian
  dynamics simulations of model hard-sphere suspensions}.  \jt{J.~Non-Newtonian
  Fluid Mech.}  \bvol{46}~(1),  \pg{1--28}.

\bibitem[Montero~de Hijes {\em et~al.\/}(2020{\natexlab{{\em a\/}}})Montero~de
  Hijes, Espinosa, Bianco, Sanz \& Vega]{montero2020interfacial}
{\sc \au{Montero~de Hijes, Pablo}, \au{Espinosa, Jorge~R}, \au{Bianco,
  Valentino}, \au{Sanz, Eduardo} \& \au{Vega, Carlos}} \yr{2020{\natexlab{{\em
  a\/}}}}  \at{Interfacial free energy and tolman length of curved
  liquid--solid interfaces from equilibrium studies}.  \jt{The Journal of
  Physical Chemistry C}  \bvol{124}~(16),  \pg{8795--8805}.

\bibitem[Montero~de Hijes {\em et~al.\/}(2020{\natexlab{{\em b\/}}})Montero~de
  Hijes, Shi, Noya, Santiso, Gubbins, Sanz \& Vega]{montero2020young}
{\sc \au{Montero~de Hijes, P}, \au{Shi, K}, \au{Noya, Eva~G}, \au{Santiso, EE},
  \au{Gubbins, KE}, \au{Sanz, E} \& \au{Vega, C}} \yr{2020{\natexlab{{\em
  b\/}}}}  \at{The young--laplace equation for a solid--liquid interface}.
  \jt{The Journal of Chemical Physics}  \bvol{153}~(19),  \pg{191102}.

\bibitem[Hoover {\em et~al.\/}(1971)Hoover, Gray \&
  Johnson]{hoover1971thermodynamic}
{\sc \au{Hoover, William~G.}, \au{Gray, Steven~G.} \& \au{Johnson, Keith~W.}}
  \yr{1971}  \at{{Thermodynamic properties of the fluid and solid phases for
  inverse power potentials}}.  \jt{The Journal of Chemical Physics}
  \bvol{55}~(3),  \pg{1128--1136}.

\bibitem[Hoover \& Ree(1968)]{hr-68}
{\sc \au{Hoover, William~G.} \& \au{Ree, Francis~H.}} \yr{1968}  \at{Melting
  transition and communal entropy for hard spheres}.  \jt{J. Chem. Phys.}
  \bvol{49}~(8),  \pg{3609--3617},  \arxiv{arXiv:
  https://doi.org/10.1063/1.1670641}.

\bibitem[Hopkins {\em et~al.\/}(2011)Hopkins, Jiao, Stillinger \&
  Torquato]{hopkins2011phase}
{\sc \au{Hopkins, Adam~B}, \au{Jiao, Yang}, \au{Stillinger, Frank~H} \&
  \au{Torquato, Salvatore}} \yr{2011}  \at{Phase diagram and structural
  diversity of the densest binary sphere packings}.  \jt{Physical Review
  Letters}  \bvol{107}~(12),  \pg{125501}.

\bibitem[Isobe \& Krauth(2015)]{isobe2015hard}
{\sc \au{Isobe, Masaharu} \& \au{Krauth, Werner}} \yr{2015}  \at{Hard-sphere
  melting and crystallization with event-chain monte carlo}.  \jt{The Journal
  of Chemical Physics}  \bvol{143}~(8),  \pg{084509}.

\bibitem[Jiao \& Torquato(2011)]{jiao2011communication}
{\sc \au{Jiao, Yang} \& \au{Torquato, Salvatore}} \yr{2011}  \at{Communication:
  A packing of truncated tetrahedra that nearly fills all of space and its
  melting properties}.  \jt{The Journal of Chemical Physics}  \bvol{135}~(15),
  \pg{151101}.

\bibitem[Johnson {\em et~al.\/}(2018)Johnson, Landrum \& Zia]{jlz-18}
{\sc \au{Johnson, Lilian~C.}, \au{Landrum, Benjamin~J.} \& \au{Zia,
  Roseanna~N.}} \yr{2018}  \at{Yield of reversible colloidal gels during flow
  start-up: release from kinetic arrest}.  \jt{Soft Matter}  \bvol{14},
  \pg{5048--5068}.

\bibitem[Johnson \& Zia(2021)]{johnson2021phase}
{\sc \au{Johnson, Lilian~C.} \& \au{Zia, Roseanna~N.}} \yr{2021}  \at{{Phase
  mechanics of colloidal gels: osmotic pressure drives non-equilibrium phase
  separation}}.  \jt{Soft Matter}  \bvol{17}~(14),  \pg{3784--3797}.

\bibitem[Johnson {\em et~al.\/}(2019)Johnson, Zia, Moghimi \&
  Petekidis]{johnson2019influence}
{\sc \au{Johnson, Lilian~C.}, \au{Zia, Roseanna~N.}, \au{Moghimi, Esmaeel} \&
  \au{Petekidis, George}} \yr{2019}  \at{{Influence of structure on the linear
  response rheology of colloidal gels}}.  \jt{J. Rheol. (N. Y. N. Y).}
  \bvol{63}~(4),  \pg{583}.

\bibitem[Kallus \& Elser(2011)]{kallus2011dense}
{\sc \au{Kallus, Yoav} \& \au{Elser, Veit}} \yr{2011}  \at{Dense-packing
  crystal structures of physical tetrahedra}.  \jt{Physical Review
  E—Statistical, Nonlinear, and Soft Matter Physics}  \bvol{83}~(3),
  \pg{036703}.

\bibitem[Karas {\em et~al.\/}(2019)Karas, Dshemuchadse, van Anders \&
  Glotzer]{karas2019phase}
{\sc \au{Karas, Andrew~S}, \au{Dshemuchadse, Julia}, \au{van Anders, Greg} \&
  \au{Glotzer, Sharon~C}} \yr{2019}  \at{Phase behavior and design rules for
  plastic colloidal crystals of hard polyhedra via consideration of directional
  entropic forces}.  \jt{Soft Matter}  \bvol{15}~(27),  \pg{5380--5389}.

\bibitem[Khair \& Brady(2006)]{kb-06}
{\sc \au{Khair, Aditya~S.} \& \au{Brady, John~F.}} \yr{2006}  \at{Single
  particle motion in colloidal dispersions: a simple model for active and
  nonlinear microrheology}.  \jt{J. Fluid Mech.}  \bvol{557},  \pg{73–117}.

\bibitem[Khair {\em et~al.\/}(2006)Khair, Swaroop \& Brady]{ksb-06}
{\sc \au{Khair, Aditya~S.}, \au{Swaroop, Manuj} \& \au{Brady, John~F.}}
  \yr{2006}  \at{A new resistance function for two rigid spheres in a uniform
  compressible low-reynolds-number flow}.  \jt{Phys. Fluids}  \bvol{18}~(4),
  \pg{043102},  \arxiv{arXiv: https://doi.org/10.1063/1.2194559}.

\bibitem[Kirkwood \& Monroe(1941)]{kirkwood1941statistical}
{\sc \au{Kirkwood, John~G} \& \au{Monroe, Elizabeth}} \yr{1941}
  \at{Statistical mechanics of fusion}.  \jt{The Journal of Chemical Physics}
  \bvol{9}~(7),  \pg{514--526}.

\bibitem[Kolafa {\em et~al.\/}(2004)Kolafa, Lab{\'\i}k \&
  Malijevsk{\`y}]{kolafa2004accurate}
{\sc \au{Kolafa, Ji{\v{r}}{\'\i}}, \au{Lab{\'\i}k, Stanislav} \&
  \au{Malijevsk{\`y}, Anatol}} \yr{2004}  \at{Accurate equation of state of the
  hard sphere fluid in stable and metastable regions}.  \jt{Physical Chemistry
  Chemical Physics}  \bvol{6}~(9),  \pg{2335--2340}.

\bibitem[Koshoji {\em et~al.\/}(2021)Koshoji, Kawamura, Fukuda \&
  Ozaki]{koshoji2021diverse}
{\sc \au{Koshoji, Ryotaro}, \au{Kawamura, Mitsuaki}, \au{Fukuda, Masahiro} \&
  \au{Ozaki, Taisuke}} \yr{2021}  \at{Diverse densest binary sphere packings
  and phase diagram}.  \jt{Physical Review E}  \bvol{103}~(2),  \pg{023307}.

\bibitem[Koshoji \& Ozaki(2021)]{koshoji2021densest}
{\sc \au{Koshoji, Ryotaro} \& \au{Ozaki, Taisuke}} \yr{2021}  \at{Densest
  ternary sphere packings}.  \jt{Physical Review E}  \bvol{104}~(2),
  \pg{024101}.

\bibitem[Kovacs(1964)]{kovacs-64}
{\sc \au{Kovacs, A.~J.}} \yr{1964}  \at{Transition vitreuse dans les
  polym{`e}res amorphes. etude ph'{e}nomenologique}.  \bt{In {\em Fortschritte
  Der Hochpolymeren-Forschung\/}}, chap.~3.  \publ{Berlin Heidelberg:
  Springer}.

\bibitem[Kranendonk \& Frenkel(1991)]{kranendonk1991computer}
{\sc \au{Kranendonk, WGT} \& \au{Frenkel, D}} \yr{1991}  \at{Computer
  simulation of solid-liquid coexistence in binary hard sphere mixtures}.
  \jt{Molecular physics}  \bvol{72}~(3),  \pg{679--697}.

\bibitem[Kratzer \& Arnold(2015)]{Kratzer2015}
{\sc \au{Kratzer, K.} \& \au{Arnold, A.}} \yr{2015}  \at{Two-stage
  crystallization of charged colloids under low supersaturation conditions}.
  \jt{Soft Matter}  \bvol{11},  \pg{2174--2182}.

\bibitem[Ladd \& Woodcock(1977)]{ladd1977triple}
{\sc \au{Ladd, AJC} \& \au{Woodcock, LV}} \yr{1977}  \at{Triple-point
  coexistence properties of the lennard-jones system}.  \jt{Chemical Physics
  Letters}  \bvol{51}~(1),  \pg{155--159}.

\bibitem[Landrum {\em et~al.\/}(2016)Landrum, Russel \& Zia]{lrz-16}
{\sc \au{Landrum, Benjamin~J.}, \au{Russel, William~B.} \& \au{Zia,
  Roseanna~N.}} \yr{2016}  \at{Delayed yield in colloidal gels: Creep, flow,
  and re-entrant solid regimes}.  \jt{J. Rheol.}  \bvol{60}~(4),
  \pg{783--807},  \arxiv{arXiv: https://doi.org/10.1122/1.4954640}.

\bibitem[Laurati {\em et~al.\/}(2005)Laurati, Stellbrink, Lund, Willner,
  Richter \& Zaccarelli]{Laurati2005}
{\sc \au{Laurati, Marco}, \au{Stellbrink, J}, \au{Lund, R}, \au{Willner, L},
  \au{Richter, D} \& \au{Zaccarelli, E}} \yr{2005}  \at{Starlike micelles with
  starlike interactions: A quantitative evaluation of structure factors and
  phase diagram}.  \jt{Physical Review Letters}  \bvol{94}~(19),  \pg{195504}.

\bibitem[Lechner \& Dellago(2008)]{Lechner2008}
{\sc \au{Lechner, W.} \& \au{Dellago, C.}} \yr{2008}  \at{Accurate
  determination of crystal structures based on averaged local bond order
  parameters}.  \jt{J. Chem. Phys}  \bvol{129},  \pg{114707}.

\bibitem[Likos(2001)]{likos2001effective}
{\sc \au{Likos, Christos~N}} \yr{2001}  \at{Effective interactions in soft
  condensed matter physics}.  \jt{Physics Reports}  \bvol{348}~(4-5),
  \pg{267--439}.

\bibitem[Likos(2006)]{likos2006soft}
{\sc \au{Likos, Christos~N}} \yr{2006}  \at{Soft matter with soft particles}.
  \jt{Soft matter}  \bvol{2}~(6),  \pg{478--498}.

\bibitem[Lim {\em et~al.\/}(2023)Lim, Lee \& Glotzer]{lim2023engineering}
{\sc \au{Lim, Yein}, \au{Lee, Sangmin} \& \au{Glotzer, Sharon~C}} \yr{2023}
  \at{Engineering the thermodynamic stability and metastability of mesophases
  of colloidal bipyramids through shape entropy}.  \jt{ACS nano}
  \bvol{17}~(5),  \pg{4287--4295}.

\bibitem[L{\"o}wen {\em et~al.\/}(1993)L{\"o}wen, Palberg \& Simon]{Lowen1993}
{\sc \au{L{\"o}wen, Hartmut}, \au{Palberg, Thomas} \& \au{Simon, Rolf}}
  \yr{1993}  \at{Dynamical criterion for freezing of colloidal liquids}.
  \jt{Physical Review Letters}  \bvol{70}~(10),  \pg{1557}.

\bibitem[Lowen \& Szamel(1993)]{Lowen1993b}
{\sc \au{Lowen, Hartmut} \& \au{Szamel, Grzegorz}} \yr{1993}  \at{Long-time
  self-diffusion coefficient in colloidal suspensions: theory versus
  simulation}.  \jt{Journal of Physics: Condensed Matter}  \bvol{5}~(15),
  \pg{2295}.

\bibitem[Marechal {\em et~al.\/}(2012)Marechal, Zimmermann \&
  L{\"o}wen]{marechal2012freezing}
{\sc \au{Marechal, Matthieu}, \au{Zimmermann, Urs} \& \au{L{\"o}wen, Hartmut}}
  \yr{2012}  \at{Freezing of parallel hard cubes with rounded edges}.  \jt{The
  Journal of Chemical Physics}  \bvol{136}~(14),  \pg{144506}.

\bibitem[McKenna(2020)]{mckenna2020looking}
{\sc \au{McKenna, Gregory~B}} \yr{2020}  \at{Looking at the glass transition:
  Challenges of extreme time scales and other interesting problems}.
  \jt{Rubber Chemistry and Technology}  \bvol{93}~(1),  \pg{79--120}.

\bibitem[McQuarrie(1975)]{mcquarrie-76}
{\sc \au{McQuarrie, Donald~A.}} \yr{1975} {\em Statistical mechanics\/}.
  \publ{Harper and Row New York}.

\bibitem[Meijer \& Frenkel(1991)]{Meijer1991}
{\sc \au{Meijer, Evert~Jan} \& \au{Frenkel, Daan}} \yr{1991}  \at{Melting line
  of yukawa system by computer simulation}.  \jt{The Journal of Chemical
  Physics}  \bvol{94}~(3),  \pg{2269--2271}.

\bibitem[Miller {\em et~al.\/}(2010)Miller, Bozorgui \&
  Cacciuto]{miller2010crystallization}
{\sc \au{Miller, William~L}, \au{Bozorgui, Behnaz} \& \au{Cacciuto, Angelo}}
  \yr{2010}  \at{Crystallization of hard aspherical particles}.  \jt{The
  Journal of Chemical Physics}  \bvol{132}~(13),  \pg{134901}.

\bibitem[Mladek {\em et~al.\/}(2007{\natexlab{{\em a\/}}})Mladek, Charbonneau
  \& Frenkel]{mladek2007phase}
{\sc \au{Mladek, Bianca~M.}, \au{Charbonneau, Patrick} \& \au{Frenkel, Daan}}
  \yr{2007{\natexlab{{\em a\/}}}}  \at{{Phase coexistence of cluster crystals:
  Beyond the gibbs phase rule}}.  \jt{Physical Review Letters}  \bvol{99}~(23),
   \pg{1--4}.

\bibitem[Mladek {\em et~al.\/}(2007{\natexlab{{\em b\/}}})Mladek, Gottwaldy,
  Kahl, Neumann \& Likos]{mladek2007clustering}
{\sc \au{Mladek, Bianca~M.}, \au{Gottwaldy, Dieter}, \au{Kahl, Gerhard},
  \au{Neumann, Martin} \& \au{Likos, Christos~N.}} \yr{2007{\natexlab{{\em
  b\/}}}}  \at{{Clustering in the absence of attractions: Density functional
  theory and computer simulations}}.  \jt{Journal of Physical Chemistry B}
  \bvol{111}~(44),  \pg{12799--12808}.

\bibitem[Moir {\em et~al.\/}(2021)Moir, Lue \& Bannerman]{moir2021tethered}
{\sc \au{Moir, Craig}, \au{Lue, Leo} \& \au{Bannerman, Marcus~N}} \yr{2021}
  \at{Tethered-particle model: The calculation of free energies for hard-sphere
  systems}.  \jt{The Journal of Chemical Physics}  \bvol{155}~(6),
  \pg{064504}.

\bibitem[Nayhouse {\em et~al.\/}(2011)Nayhouse, Amlani \&
  Orkoulas]{nayhouse2011monte}
{\sc \au{Nayhouse, Michael}, \au{Amlani, Ankur~M} \& \au{Orkoulas, G}}
  \yr{2011}  \at{A monte carlo study of the freezing transition of hard
  spheres}.  \jt{Journal of Physics: Condensed Matter}  \bvol{23}~(32),
  \pg{325106}.

\bibitem[N{\'e}meth \& Likos(1995)]{nemeth1995solid}
{\sc \au{N{\'e}meth, Zs~T} \& \au{Likos, CN}} \yr{1995}  \at{Solid to solid
  isostructural transitions: the case of attractive yukawa potentials}.
  \jt{Journal of Physics: Condensed Matter}  \bvol{7}~(41),  \pg{L537}.

\bibitem[Noya {\em et~al.\/}(2008)Noya, Vega \&
  de~Miguel]{noya2008determination}
{\sc \au{Noya, Eva~G}, \au{Vega, Carlos} \& \au{de~Miguel, Enrique}} \yr{2008}
  \at{Determination of the melting point of hard spheres from direct
  coexistence simulation methods}.  \jt{The Journal of Chemical Physics}
  \bvol{128}~(15),  \pg{154507}.

\bibitem[Odriozola(2009)]{odriozola2009replica}
{\sc \au{Odriozola, Gerardo}} \yr{2009}  \at{Replica exchange monte carlo
  applied to hard spheres}.  \jt{The Journal of Chemical Physics}
  \bvol{131}~(14),  \pg{144107}.

\bibitem[Onsager(1949)]{onsager1949effects}
{\sc \au{Onsager, Lars}} \yr{1949}  \at{The effects of shape on the interaction
  of colloidal particles}.  \jt{Annals of the New York Academy of Sciences}
  \bvol{51}~(4),  \pg{627--659}.

\bibitem[Padmanabhan \& Zia(2018)]{pz-18}
{\sc \au{Padmanabhan, Poornima} \& \au{Zia, Roseanna}} \yr{2018}
  \at{Gravitational collapse of colloidal gels: non-equilibrium phase
  separation driven by osmotic pressure}.  \jt{Soft Matter}  \bvol{14},
  \pg{3265--3287}.

\bibitem[Pelaez-Fernandez {\em et~al.\/}(2015)Pelaez-Fernandez, Souslov, Lyon,
  Goldbart \& Fernandez-Nieves]{pelaez2015impact}
{\sc \au{Pelaez-Fernandez, M.}, \au{Souslov, Anton}, \au{Lyon, L.~A.},
  \au{Goldbart, P.~M.} \& \au{Fernandez-Nieves, A.}} \yr{2015}  \at{{Impact of
  single-particle compressibility on the fluid-solid phase transition for ionic
  microgel suspensions}}.  \jt{Physical Review Letters}  \bvol{114}~(9),
  \pg{2--6}.

\bibitem[Peng \& McKenna(2016)]{pm-16}
{\sc \au{Peng, Xiaoguang} \& \au{McKenna, Gregory~B.}} \yr{2016}  \at{Physical
  aging and structural recovery in a colloidal glass subjected to
  volume-fraction jump conditions}.  \jt{Phys. Rev. E}  \bvol{93},
  \pg{042603}.

\bibitem[Peroukidis \& Vanakaras(2013)]{peroukidis2013phase}
{\sc \au{Peroukidis, Stavros~D} \& \au{Vanakaras, Alexandros~G}} \yr{2013}
  \at{Phase diagram of hard board-like colloids from computer simulations}.
  \jt{Soft Matter}  \bvol{9}~(31),  \pg{7419--7423}.

\bibitem[Phan {\em et~al.\/}(1996)Phan, Russel, Cheng, Zhu, Chaikin, Dunsmuir
  \& Ottewill]{prczcdo-96}
{\sc \au{Phan, See-Eng}, \au{Russel, William}, \au{Cheng, Zhengdong}, \au{Zhu,
  Jixiang}, \au{Chaikin, Paul}, \au{Dunsmuir, John} \& \au{Ottewill, Ronald}}
  \yr{1996}  \at{{Phase transition, equation of state, and limiting shear
  viscosities of hard sphere dispersions}}.  \jt{Phys. Rev. E}  \bvol{54}~(6),
  \pg{6633--6645}.

\bibitem[Piazza {\em et~al.\/}(1993)Piazza, Bellini \&
  Degiorgio]{piazza1993equilibrium}
{\sc \au{Piazza, Roberto}, \au{Bellini, Tommaso} \& \au{Degiorgio, Vittorio}}
  \yr{1993}  \at{Equilibrium sedimentation profiles of screened charged
  colloids: A test of the hard-sphere equation of state}.  \jt{Physical Review
  Letters}  \bvol{71}~(25),  \pg{4267}.

\bibitem[Pieprzyk {\em et~al.\/}(2019)Pieprzyk, Bannerman, Bra{\'n}ka, Chudak
  \& Heyes]{pieprzyk2019thermodynamic}
{\sc \au{Pieprzyk, S{\l}awomir}, \au{Bannerman, Marcus~N}, \au{Bra{\'n}ka,
  Arkadiusz~C}, \au{Chudak, Maciej} \& \au{Heyes, David~M}} \yr{2019}
  \at{Thermodynamic and dynamical properties of the hard sphere system
  revisited by molecular dynamics simulation}.  \jt{Physical Chemistry Chemical
  Physics}  \bvol{21}~(13),  \pg{6886--6899}.

\bibitem[Poon {\em et~al.\/}(2012)Poon, Weeks \& Royall]{poon2012measuring}
{\sc \au{Poon, Wilson C.~K.}, \au{Weeks, Eric~R.} \& \au{Royall, C.~Patrick}}
  \yr{2012}  \at{On measuring colloidal volume fractions}.  \jt{Soft matter}
  \bvol{8}~(1),  \pg{21--30}.

\bibitem[Pusey \& van Megen(1986)]{pvM-86}
{\sc \au{Pusey, P.~N.} \& \au{van Megen, W.}} \yr{1986}  \at{{Phase behavior of
  concentrated suspensions of nearly hard colloidal spheres}}.  \jt{Nature}
  \bvol{320},  \pg{340--342}.

\bibitem[Ree \& Hoover(1964)]{ree1964fifth}
{\sc \au{Ree, Francis~H} \& \au{Hoover, William~G}} \yr{1964}  \at{Fifth and
  sixth virial coefficients for hard spheres and hard disks}.  \jt{The Journal
  of Chemical Physics}  \bvol{40}~(4),  \pg{939--950}.

\bibitem[Ree \& Hoover(1967)]{ree1967seventh}
{\sc \au{Ree, Francis~H} \& \au{Hoover, William~G}} \yr{1967}  \at{Seventh
  virial coefficients for hard spheres and hard disks}.  \jt{The Journal of
  Chemical Physics}  \bvol{46}~(11),  \pg{4181--4197}.

\bibitem[Robbins {\em et~al.\/}(1988)Robbins, Kremer \& Grest]{Robbins1988}
{\sc \au{Robbins, Mark~O}, \au{Kremer, Kurt} \& \au{Grest, Gary~S}} \yr{1988}
  \at{Phase diagram and dynamics of yukawa systems}.  \jt{The Journal of
  Chemical Physics}  \bvol{88}~(5),  \pg{3286--3312}.

\bibitem[Royall {\em et~al.\/}(2024)Royall, Charbonneau, Dijkstra, Russo,
  Smallenburg, Speck \& Valeriani]{royall2024colloidal}
{\sc \au{Royall, C~Patrick}, \au{Charbonneau, Patrick}, \au{Dijkstra,
  Marjolein}, \au{Russo, John}, \au{Smallenburg, Frank}, \au{Speck, Thomas} \&
  \au{Valeriani, Chantal}} \yr{2024}  \at{Colloidal hard spheres: Triumphs,
  challenges, and mysteries}.  \jt{Reviews of Modern Physics}  \bvol{96}~(4),
  \pg{045003}.

\bibitem[Royall {\em et~al.\/}(2013)Royall, Poon \& Weeks]{rpw-13}
{\sc \au{Royall, C.~Patrick}, \au{Poon, Wilson C.~K.} \& \au{Weeks, Eric~R.}}
  \yr{2013}  \at{In search of colloidal hard spheres}.  \jt{Soft Matter}
  \bvol{9},  \pg{17--27}.

\bibitem[Russel {\em et~al.\/}(1991)Russel, Russel, Saville \&
  Schowalter]{russel1991colloidal}
{\sc \au{Russel, William~Bailey}, \au{Russel, WB}, \au{Saville, Dudley~A} \&
  \au{Schowalter, William~Raymond}} \yr{1991} {\em Colloidal dispersions\/}.
  \publ{Cambridge university press}.

\bibitem[Rutgers {\em et~al.\/}(1996)Rutgers, Dunsmuir, Xue, Russel \&
  Chaikin]{rutgers1996measurement}
{\sc \au{Rutgers, MA}, \au{Dunsmuir, JH}, \au{Xue, J-Z}, \au{Russel, WB} \&
  \au{Chaikin, PM}} \yr{1996}  \at{Measurement of the hard-sphere equation of
  state using screened charged polystyrene colloids}.  \jt{Physical Review B}
  \bvol{53}~(9),  \pg{5043}.

\bibitem[Ryu {\em et~al.\/}(2022)Ryu, Fenton, Nguyen, Helgeson \&
  Zia]{ryu2022modeling}
{\sc \au{Ryu, Brian~K.}, \au{Fenton, Scott~M.}, \au{Nguyen, Tuan~T.D.},
  \au{Helgeson, Matthew~E.} \& \au{Zia, Roseanna~N.}} \yr{2022}  \at{{Modeling
  colloidal interactions that predict equilibrium and non-equilibrium states}}.
   \jt{J. Chem. Phys.}  \bvol{156}~(22),  \pg{224101}.

\bibitem[Sanchez-Burgos {\em et~al.\/}(2021)Sanchez-Burgos, Sanz, Vega \&
  Espinosa]{sanchez2021fcc}
{\sc \au{Sanchez-Burgos, Ignacio}, \au{Sanz, Eduardo}, \au{Vega, Carlos} \&
  \au{Espinosa, Jorge~R}} \yr{2021}  \at{Fcc vs. hcp competition in colloidal
  hard-sphere nucleation: on their relative stability, interfacial free energy
  and nucleation rate}.  \jt{Physical Chemistry Chemical Physics}
  \bvol{23}~(35),  \pg{19611--19626}.

\bibitem[Schultz \& Kofke(2014)]{schultz2014fifth}
{\sc \au{Schultz, Andrew~J} \& \au{Kofke, David~A}} \yr{2014}  \at{Fifth to
  eleventh virial coefficients of hard spheres}.  \jt{Physical Review E}
  \bvol{90}~(2),  \pg{023301}.

\bibitem[Senff \& Richtering(1999)]{senff1999temperature}
{\sc \au{Senff, H.} \& \au{Richtering, W.}} \yr{1999}  \at{{Temperature
  sensitive microgel suspensions: Colloidal phase behavior and rheology of soft
  spheres}}.  \jt{Journal of Chemical Physics}  \bvol{111}~(4),
  \pg{1705--1711}.

\bibitem[Song {\em et~al.\/}(2022)Song, Smith, Zhu, Adams, Colby, Finnegan,
  Gough, Hillery, Irvine, Maji \& St.~John]{Anvil}
{\sc \au{Song, X.~Carol}, \au{Smith, Preston nd~Kalyanam, Rajesh}, \au{Zhu,
  Xiao}, \au{Adams, Eric}, \au{Colby, Kevin}, \au{Finnegan, Patrick},
  \au{Gough, Erik}, \au{Hillery, Elizabett}, \au{Irvine, Rick}, \au{Maji,
  Amiya} \& \au{St.~John, Jason}} \yr{2022} Anvil - system architecture and
  experiences from deployment and early user operations.  \bt{In {\em Practice
  and Experience in Advanced Research Computing 2022: Revolutionary: Computing,
  Connections, You\/}}, {\em PEARC '22\/} .  \publ{New York, NY, USA:
  Association for Computing Machinery}.

\bibitem[Speedy(1997)]{speedy1997pressure}
{\sc \au{Speedy, Robin~J}} \yr{1997}  \at{Pressure of the metastable
  hard-sphere fluid}.  \jt{Journal of Physics: Condensed Matter}
  \bvol{9}~(41),  \pg{8591}.

\bibitem[Statt {\em et~al.\/}(2016)Statt, Schmitz, Virnau \&
  Binder]{statt2016monte}
{\sc \au{Statt, Antonia}, \au{Schmitz, Fabian}, \au{Virnau, Peter} \&
  \au{Binder, Kurt}} \yr{2016} Monte carlo simulation of crystal-liquid phase
  coexistence.  \bt{In {\em High Performance Computing in Science and
  Engineering{\'{}} 15: Transactions of the High Performance Computing Center,
  Stuttgart (HLRS) 2015\/}},  \pg{pp. 75--87}. Springer.

\bibitem[Steinhardt {\em et~al.\/}(1983)Steinhardt, Nelson \&
  Ronchetti]{Steinhardt1983}
{\sc \au{Steinhardt, P.~J.}, \au{Nelson, D.~R.} \& \au{Ronchetti, M.}}
  \yr{1983}  \at{Bond-orientational order in liquids and glassess}.  \jt{Phys.
  Rev. B, PRB}  \bvol{28}~(2),  \pg{784--805}.

\bibitem[Sunol \& Zia(2023)]{sunol2023confined}
{\sc \au{Sunol, Alp~M.} \& \au{Zia, Roseanna~N.}} \yr{2023}  \at{{Confined
  Brownian suspensions: Equilibrium diffusion, thermodynamics, and rheology}}.
  \jt{Journal of Rheology}  \bvol{67}~(2),  \pg{433--460},  \arxiv{arXiv:
  https://pubs.aip.org/sor/jor/article-pdf/67/2/433/19824761/433\_1\_online.pdf}.

\bibitem[Swaroop \& Brady(2007)]{sb-07}
{\sc \au{Swaroop, Manuj} \& \au{Brady, John~F.}} \yr{2007}  \at{The bulk
  viscosity of suspensions}.  \jt{J. Rheol.}  \bvol{51}~(3),  \pg{409--428},
  \arxiv{arXiv: https://doi.org/10.1122/1.2714643}.

\bibitem[Tateno {\em et~al.\/}(2019)Tateno, Yanagishima, Russo \&
  Tanaka]{tateno2019influence}
{\sc \au{Tateno, Michio}, \au{Yanagishima, Taiki}, \au{Russo, John} \&
  \au{Tanaka, Hajime}} \yr{2019}  \at{Influence of hydrodynamic interactions on
  colloidal crystallization}.  \jt{Phys. Rev. Lett.}  \bvol{123},  \pg{258002}.

\bibitem[Thiele(1963)]{thiele1963equation}
{\sc \au{Thiele, Everett}} \yr{1963}  \at{Equation of state for hard spheres}.
  \jt{The Journal of Chemical Physics}  \bvol{39}~(2),  \pg{474--479}.

\bibitem[Thompson {\em et~al.\/}(2022)Thompson, Aktulga, Berger, Bolintineanu,
  Brown, Crozier, in~'t Veld, Kohlmeyer, Moore, Nguyen, Shan, Stevens,
  Tranchida, Trott \& Plimpton]{thompson2022lammps}
{\sc \au{Thompson, A.~P.}, \au{Aktulga, H.~M.}, \au{Berger, R.},
  \au{Bolintineanu, D.~S.}, \au{Brown, W.~M.}, \au{Crozier, P.~S.}, \au{in~'t
  Veld, P.~J.}, \au{Kohlmeyer, A.}, \au{Moore, S.~G.}, \au{Nguyen, T.~D.},
  \au{Shan, R.}, \au{Stevens, M.~J.}, \au{Tranchida, J.}, \au{Trott, C.} \&
  \au{Plimpton, S.~J.}} \yr{2022}  \at{{LAMMPS} - a flexible simulation tool
  for particle-based materials modeling at the atomic, meso, and continuum
  scales}.  \jt{Comp. Phys. Comm.}  \bvol{271},  \pg{108171}.

\bibitem[Uhlenbeck(1963)]{uhlenbeck1963p}
{\sc \au{Uhlenbeck, GE}} \yr{1963}  \at{The many-body problem}.  \bt{In {\em
  Proceedings of the Symposium on the Many-body Problem Held at Stevens
  Institute of Technology, Hoboken, New Jersey, January 28-29, 1957\/} (ed.
  \ed{JK~Percus})},  \pg{p. 498}.  \publ{Interscience Publishers/John Wiley,
  London}.

\bibitem[Ustinov(2017)]{ustinov2017thermodynamics}
{\sc \au{Ustinov, EA}} \yr{2017}  \at{Thermodynamics and simulation of
  hard-sphere fluid and solid: Kinetic monte carlo method versus standard
  metropolis scheme}.  \jt{The Journal of Chemical Physics}  \bvol{146}~(3),
  \pg{034110}.

\bibitem[Vega \& Noya(2007)]{vega2007revisiting}
{\sc \au{Vega, Carlos} \& \au{Noya, Eva~G}} \yr{2007}  \at{Revisiting the
  frenkel-ladd method to compute the free energy of solids: The einstein
  molecule approach}.  \jt{The Journal of Chemical Physics}  \bvol{127}~(15),
  \pg{154113}.

\bibitem[Vlassopoulos \& Cloitre(2014)]{Vlassopoulos2014}
{\sc \au{Vlassopoulos, Dimitris} \& \au{Cloitre, Michel}} \yr{2014}
  \at{Tunable rheology of dense soft deformable colloids}.  \jt{Current opinion
  in colloid \& interface science}  \bvol{19}~(6),  \pg{561--574}.

\bibitem[Volmer \& Weber(1926)]{CNT1926}
{\sc \au{Volmer, M.} \& \au{Weber, A.}} \yr{1926}  \at{{Keimbildung in
  \"{u}bers\"{a}ttigten Gebilden}}.  \jt{Zeitschrift f\"{u}r Physikalische
  Chemie}  \bvol{119U}~(1),  \pg{277--301}.

\bibitem[Wang {\em et~al.\/}(2018)Wang, Wang, Peng, Zheng \&
  Han]{wang2018homogeneous}
{\sc \au{Wang, Feng}, \au{Wang, Ziren}, \au{Peng, Yi}, \au{Zheng, Zhongyu} \&
  \au{Han, Yilong}} \yr{2018}  \at{Homogeneous melting near the superheat limit
  of hard-sphere crystals}.  \jt{Soft Matter}  \bvol{14}~(13),
  \pg{2447--2453}.

\bibitem[Wang {\em et~al.\/}(in review)Wang, Dhumal, Zakhari \&
  Zia]{wangInreviewElusive}
{\sc \au{Wang, J.~Galen}, \au{Dhumal, Umesh}, \au{Zakhari, Monica E.~A.} \&
  \au{Zia, Roseanna~N.}} \yr{in review}  \at{The elusive liquid-and-crystal
  coexistence state in simulations of monodisperse, hard-sphere colloids}.
  \jt{Submitted to AIChE Journal} .

\bibitem[Wang \& Zia(2021)]{wang2021vitrification}
{\sc \au{Wang, J~Galen} \& \au{Zia, Roseanna~N}} \yr{2021}  \at{Vitrification
  is a spontaneous non-equilibrium transition driven by osmotic pressure}.
  \jt{Journal of Physics: Condensed Matter}  \bvol{33}~(18),  \pg{184002}.

\bibitem[Wertheim(1963)]{wertheim1963exact}
{\sc \au{Wertheim, MS}} \yr{1963}  \at{Exact solution of the percus-yevick
  integral equation for hard spheres}.  \jt{Physical Review Letters}
  \bvol{10}~(8),  \pg{321}.

\bibitem[Wilding \& Bruce(2000)]{wilding2000freezing}
{\sc \au{Wilding, NB} \& \au{Bruce, AD}} \yr{2000}  \at{Freezing by monte carlo
  phase switch}.  \jt{Physical Review Letters}  \bvol{85}~(24),  \pg{5138}.

\bibitem[Wilding \& Sollich(2010)]{wilding2010phase}
{\sc \au{Wilding, Nigel~B} \& \au{Sollich, Peter}} \yr{2010}  \at{Phase
  behavior of polydisperse spheres: Simulation strategies and an application to
  the freezing transition}.  \jt{The Journal of Chemical Physics}
  \bvol{133}~(22),  \pg{224102}.

\bibitem[W{\"o}hler \& Schilling(2022)]{wohler2022hard}
{\sc \au{W{\"o}hler, Wilkin} \& \au{Schilling, Tanja}} \yr{2022}  \at{Hard
  sphere crystal nucleation rates: Reconciliation of simulation and
  experiment}.  \jt{Physical Review Letters}  \bvol{128}~(23),  \pg{238001}.

\bibitem[ten Wolde {\em et~al.\/}(1996)ten Wolde, Ruiz-Montero \&
  Frenkel]{tenwolde1996numerical}
{\sc \au{ten Wolde, Pieter~R}, \au{Ruiz-Montero, Maria~J} \& \au{Frenkel,
  Daan}} \yr{1996}  \at{Numerical calculation of the rate of crystal nucleation
  in a lennard-jones system at moderate undercooling}.  \jt{The Journal of
  Chemical Physics}  \bvol{104}~(24),  \pg{9932--9947}.

\bibitem[Woodcock(1997)]{Woodcock1997}
{\sc \au{Woodcock, LV}} \yr{1997}  \at{Entropy difference between the
  face-centred cubic and hexagonal close-packed crystal structures}.
  \jt{Nature}  \bvol{385}~(6612),  \pg{141--143}.

\bibitem[Zaccarelli {\em et~al.\/}(2009)Zaccarelli, Valeriani, Sanz, Poon,
  Cates \& Pusey]{zaccarelli2009crystallization}
{\sc \au{Zaccarelli, E.}, \au{Valeriani, C.}, \au{Sanz, E.}, \au{Poon, W.
  C.~K.}, \au{Cates, M.~E.} \& \au{Pusey, P.~N.}} \yr{2009}
  \at{Crystallization of hard-sphere glasses}.  \jt{Phys. Rev. Lett.}
  \bvol{103},  \pg{135704}.

\bibitem[Zakhari {\em et~al.\/}(2017)Zakhari, Anderson \&
  H{\"u}tter]{Zakhari2017}
{\sc \au{Zakhari, Monica~EA}, \au{Anderson, Patrick~D} \& \au{H{\"u}tter,
  Markus}} \yr{2017}  \at{Effect of particle-size dynamics on properties of
  dense spongy-particle systems: Approach towards equilibrium}.  \jt{Physical
  Review E}  \bvol{96}~(1),  \pg{012604}.

\bibitem[Zia \& Brady(2012)]{zb-12}
{\sc \au{Zia, R.~N.} \& \au{Brady, J.~F.}} \yr{2012}  \at{Microviscosity,
  microdiffusivity, and normal stresses in colloidal dispersions}.  \jt{J.
  Rheol.}  \bvol{56}~(5),  \pg{1175--1208},  \arxiv{arXiv:
  https://doi.org/10.1122/1.4722880}.

\bibitem[Zia {\em et~al.\/}(2014)Zia, Landrum \& Russel]{zlr-14}
{\sc \au{Zia, Roseanna~N.}, \au{Landrum, Benjamin~J.} \& \au{Russel,
  William~B.}} \yr{2014}  \at{A micro-mechanical study of coarsening and
  rheology of colloidal gels: Cage building, cage hopping, and smoluchowski's
  ratchet}.  \jt{J. Rheol.}  \bvol{58}~(5),  \pg{1121--1157},  \arxiv{arXiv:
  https://doi.org/10.1122/1.4892115}.

\bibitem[Zia {\em et~al.\/}(2015)Zia, Swan \& Su]{zss-15}
{\sc \au{Zia, Roseanna~N.}, \au{Swan, James~W.} \& \au{Su, Yu}} \yr{2015}
  \at{Pair mobility functions for rigid spheres in concentrated colloidal
  dispersions: Force, torque, translation, and rotation}.  \jt{J. Chem. Phys.}
  \bvol{143}~(22),  \pg{224901},  \arxiv{arXiv:
  https://doi.org/10.1063/1.4936664}.

\bibitem[Zubarev \& Iskakova(2005)]{zubarev2005condensation}
{\sc \au{Zubarev, A~Yu} \& \au{Iskakova, L~Yu}} \yr{2005}  \at{Condensation
  phase transitions in bidisperse colloids}.  \jt{Physica A: Statistical
  Mechanics and its Applications}  \bvol{349}~(1-2),  \pg{1--10}.

\bibitem[Zykova-Timan {\em et~al.\/}(2010)Zykova-Timan, Horbach \&
  Binder]{zykova2010monte}
{\sc \au{Zykova-Timan, Tatyana}, \au{Horbach, J{\"u}rgen} \& \au{Binder, Kurt}}
  \yr{2010}  \at{Monte carlo simulations of the solid-liquid transition in hard
  spheres and colloid-polymer mixtures}.  \jt{The Journal of Chemical Physics}
  \bvol{133}~(1),  \pg{014705}.

\end{thebibliography}
\bibliographystyle{jfm}

\end{document}